\documentclass[10pt,conference]{IEEEtran}
\usepackage{amsmath}
\usepackage{cite}
\usepackage{graphicx}
\usepackage{amsfonts}
\usepackage{latexsym}
\usepackage{subfigure}
\usepackage{algorithm}
\usepackage{algorithmic}
\usepackage{color}
\usepackage{times}
\usepackage{epsfig, graphics}
\usepackage{bm}
\usepackage{subfigure}
\usepackage{graphicx,wrapfig,lipsum}
\usepackage{amsthm}
\begin{document}
\newtheorem{theorem}{Theorem}
\newtheorem{lemma}{Lemma}
\newtheorem{conjecture}{Conjecture}
\newtheorem{corollary}{Corollary}
\newtheorem{definition}{Definition}
\newtheorem{scheme}{Scheme}
\newcommand{\argmax}{\arg\!\max}
\newcommand{\rev}[1]{{\color{red}#1}}
\newcommand{\pound}{\operatornamewithlimits{\gtrless}}
\IEEEoverridecommandlockouts

\title{Cross-layer Design and SDR Implementation of DSA, Backpressure Routing and Network Coding}
\author{\IEEEauthorblockN{Sohraab Soltani, Yalin E. Sagduyu, Sean Scanlon, Yi Shi, Jason H. Li, Jared Feldman, and John D. Matyjas}
\thanks{Sohraab Soltani, Yalin E. Sagduyu, Sean Scanlon and Jason H. Li are with Intelligent Automation, Inc., Rockville, MD, USA; Email: \{ssoltani, ysagduyu, sscanlon, jli\}@i-a-i.com. Yi Shi is with Intelligent Automation, Inc. and Virginia Tech, Blacksburg, VA, USA; Email: yshi@vt.edu. Jared Feldman, and John D. Matyjas are with U.S. Air Force Research Laboratory, RITF, Rome, NY 13441; Email: \{jared.feldman, john.matyjas\}@us.af.mil}
\thanks{DISTRIBUTION A. Approved for public release; distribution unlimited. (AFRL/RITF; 88ABW-2018-1424). Any opinions, findings and conclusions or recommendations expressed in this material are those of the authors and do not necessarily reflect the views of the Air Force.}
\thanks{Preliminary version of the material in this paper was partially presented at IEEE Military Communications (MILCOM) 2015 \cite{Soltani} and 2017 \cite{Soltani2}.}
}

\maketitle

\begin{abstract}
A cross-layer cognitive radio system is designed to support unicast and multicast traffic with integration of dynamic spectrum access (DSA), backpressure algorithm, and network coding for multi-hop networking. The full protocol stack that operates with distributed coordination and local information exchange is implemented with software-defined radios (SDRs) and assessed in a realistic test and evaluation (T\&E) system based on a network emulation testbed. Without a common control channel, each SDR performs neighborhood discovery, spectrum sensing and channel estimation, and executes a distributed extension of backpressure algorithm that optimizes the spectrum utility (that represents link rates and traffic congestion) with joint DSA and routing. The backpressure algorithm is extended to support multicast traffic with network coding deployed over virtual queues (for multicast destinations). In addition to full rank decoding at destinations, rank deficient decoding is also considered to reduce the delay. Cognitive network functionalities are programmed with GNU Radio and Python modules are developed for different layers. USRP radios are used as RF front ends. A wireless network T\&E system is presented to execute emulation tests, where radios communicate with each other through a wireless network emulator that controls physical channels according to path loss, fading, and topology effects. Emulation tests are presented for different topologies to evaluate the throughput, backlog and energy consumption. Results verify the SDR implementation and the joint effect of DSA, backpressure routing and network coding under realistic channel and radio hardware effects.
\end{abstract}

\section{Introduction}
While spectrum resources are scarce across time, space and frequency dimensions, static allocation with fixed channel assignments does not efficiently utilize the spectrum.
Cognitive radio fills this gap by
discovering new spectrum opportunities and adapting communication chain to spectrum dynamics.
\emph{Routing and spectrum access} decisions can be made jointly in a \emph{cross-layer} solution to utilize the potential correlation of spectrum occupancy in time and space \cite{Akyildiz}. With the proliferation of commercial off-the-shelf software-defined radios (SDRs), it has become feasible to deploy a multi-hop, multi-channel cognitive radio system with distributed and decentralized coordination. In this paper, we first present step by step how to design a \emph{protocol stack} for a cognitive radio network and implement it with SDR. Then we describe how to build a high fidelity \emph{test and evaluation (T\&E) system} with a network channel emulation capability, and  assess the performance of the implemented cognitive radio network solution in different network scenarios.

There have been various efforts to develop either device-level SDR solutions or isolated protocol-level solutions such as dynamic spectrum access (DSA) and spectrum sensing. In this paper, we consider the multi-hop cognitive radio network as an \emph{end-to-end system} with the full protocol stack  implementation that provides all the necessary for real system deployment including neighborhood discovery, spectrum sensing, channel estimation, channel rendezvous for distributed coordination, spectrum allocation (frequency allocation, power control and channel access), and \emph{cross-layer} design that integrates DSA, routing and network coding. This solution captures temporal and spatial spectrum opportunities, and optimizes the spectrum utility (that combines links rates with traffic congestion) by integrating \emph{backpressure algorithm} (also called backpressure routing) with \emph{network coding} to support both \emph{unicast} and \emph{multicast} traffic. We present the implementation of a \emph{distributed} protocol at each SDR that relies on local information exchange and distributed decision making without using any centralized scheduler or common control channel that would otherwise pose a communication bottleneck and a single point of failure.

The backpressure algorithm is known to achieve the optimal throughput (within the stability region) for dynamic traffic by \emph{jointly} optimizing routing and centralized scheduling decisions \cite{Tassiulas}. The distributed extension of backpressure algorithm enables dynamic routing in a spectrally diverse cognitive radio system without maintaining end-to-end paths. Backpressure algorithm has been applied to cognitive radio networks in \cite{Ding1} and its performance has been evaluated through simulation studies.
There have been  efforts to implement the backpressure algorithm on real radio devices such as on WiFi cards in a wireless testbed \cite{Laufer, Choumas} and on stand-alone WiFi radios in an emulation testbed \cite{Ding} . The default setting for the backpressure algorithm is the unicast traffic where each flow has a single source-destination pair \cite{Tassiulas,Ding1,Ding}. The backpressure algorithm can be extended to \emph{multicast traffic} by setting up and maintaining multi-hop tree structures \cite{Sarkar}. However, this poses a challenge for a distributed cognitive radio network, where routing and DSA decisions should account for fast spectrum dynamics and it is infeasible to maintain multi-hop tree routes.

To support multicast traffic in a cognitive radio network, we integrate \emph{network coding} with the backpressure algorithm. Network coding combines information flows to achieve the maximum achievable rates \cite{Ahlswede}. For the case of dynamic multicast traffic, \cite{Ho2} provided the theoretical foundation that integrates network coding with the backpressure algorithm. Random linear network coding (RLNC) provides a practical way of distributed decision making in multi-hop networks \cite{Ho1}.
Network coding was broadly applied to wireless systems through joint design with medium access control (MAC) \cite{Sagduyu2,Sagduyu6, Wu1, Sagduyu13, Sagduyu1, Sagduyu5}. To account for stochastic packet traffic, network coding was also studied in terms of queue stability and its implications on throughput and delay. These studies were typically focused on one-hop networks \cite{Sagduyu3, Sagduyu15, Sagduyu10, Sagduyu4, Sagduyu7, Sagduyu9, Yang1, Sagduyu8, Yang2} or simple extensions of them such as line networks \cite{Sagduyu11, Berry, Sagduyu12} and two-way relay networks \cite{Ertugrul1, Ertugrul2, Ertugrul3}. Network coding brings the advantage of shaping spectrum availability to a more predictable form and generates additional spectrum opportunities that can be used by cognitive radios for DSA \cite{Shanshan1,Shanshan2,Shanshan3,Fanous1,Fanous2}.

While network coding was applied to cognitive radio networks through the joint design with DSA, these studies were limited to specific wireless system configurations, e.g., \cite{Zhang} presented a cross layer approach that explores the joint advantage of network coding and dynamic subcarrier assignment, \cite{Xin} applied network coding to traffic between primary and secondary user to enable secondary users to serve as relays for primary user traffic, \cite{Jin} considered broadcast from a base station to all secondary users, and \cite{Qu} studied multicast in multi-hop cognitive radio networks in terms of flow optimization while accounting for spectrum uncertainty but excluding the effects of stochastic traffic. From the perspective of scaling laws, \cite{ZWang} analyzed the capacity and delay of cognitive radio networks with both routing and network coding.

However, a clear understanding is largely missing for practical design and real-system implementation of network coding in a \emph{general cognitive radio system}, where all stochastic traffic modes including \emph{unicast and multicast traffic} are supported in a general network topology by the joint design of network coding with DSA and backpressure algorithm.

Similar to the backpressure algorithm, network coding  was also applied to real systems such as social mobile clouds \cite{Fitzek}, device to device communications \cite{Pahlevani}, embedded platforms such as Raspberry Pi \cite{Paramanathan} and WiFi radios such as RouterStation Pros \cite{Sagduyu16, JZhang}, 802.11-based mobile ad hoc networks \cite{Firoiu1, Firoiu2}, and transmission control protocol (TCP) \cite{TCP}. However, the implementation of full-stack cross-layer protocols (including the combination of DSA, backpressure algorithm and network coding) with SDRs is missing and there is no solution implemented with SDRs while accounting for real traffic, channel, and radio hardware effects in multi-hop cognitive radio systems.

In this paper, we combine \emph{DSA, backpressure algorithm and network coding} in a \emph{cross-layer} design and implementation. We define a local spectrum utility to represent both channel and traffic effects. This utility is maximized by individual nodes running the backpressure algorithm (that is extended to allow the controlled use of cycles in network flows) in a distributed optimization framework to determine link schedules and routes along with spectrum allocation. The traditional backpressure algorithm is based on centralized scheduling. Instead, we apply a distributed coordination mechanism without centralized scheduler and common control channel to support neighborhood discovery, spectrum sensing, channel rendezvous, and local information exchange. We implement the \emph{virtual queue} system \cite{Ho2} at each node to combine network coding with the backpressure algorithm. First, we consider full rank decoding with \emph{Gaussian elimination} to decode the network-coded packets after waiting to accumulate full rank of coded packets at the destinations. Then we apply \emph{rank deficient decoding} \cite{Yan} to reduce the time to accumulate and decode the network-coded packets. To improve the decoding accuracy, our design applies precoding on ranks of network codes. We verify this design  by the implementation with USRP N210 radios \cite{Ettus} using GNU radio \cite{Blossom} and Python modules for different layers.

In current practice, wireless network evaluation is mainly performed either through software simulations or hardware testbeds. These approaches lack the fidelity and flexibility needed for accurate and realistic evaluation of a wireless network \cite{Camp}. Software simulations, such as ns-2/3, typically use simplistic physical layer modeling that ignores various effects, e.g., nonlinearity, filtering and inter modulation by hardware. There have been several developments of cognitive radio or SDR testbeds \cite{Nishra05:testbed, Kaushik}, e.g., ORBIT \cite{Raychaudhuri} and CORNET \cite{Virginia}. These testbeds can only represent static scenarios without fully controlling or reliably repeating experiments and cannot represent an arbitrary topology. Cognitive radios involve various configurable parameters. Therefore, a programmable, controllable and repeatable T\&E system is needed to assess the performance of cognitive radio network under different configurations.

To fill this gap, we built a wireless network T\&E system based on an emulation testbed, where the dynamic network topology is controlled by a high fidelity network channel emulator. Hardware-in-the-loop network emulation tests provide the flexibility to change channel effects and repeat the same tests for different network protocols under the same channel conditions \cite{Kwak1, Kwak2, Ding2, Soltani3, Sagduyu14}. RFnest is used as the network channel emulator to digitally control network connectivity, mobility and channel conditions, and facilitate repeatable testing and evaluation under identical scenarios \cite{Yackoski, Yackoski2}. In this testbed, USRP N210 radios transmit real signals to each other over emulated channels with dynamically controlled topology, mobility and channel impulse response properties.
Using four network topologies (line, grid, ring and butterfly), throughput, (queue) backlog, energy consumption and overhead are measured in emulation tests with real radios. Overhead is the number of control packets exchanged to implement the network protocols and it is different from the network coding overhead (namely, the number of coding coefficients carried along with packets) that was studied in \cite{Honig1,Honig2}. The emulations tests verify the SDR implementation and demonstrate the joint effect of DSA, backpressure algorithm and network coding in optimizing the cognitive radio system performance.

Our novel contributions are summarized as follows:
\begin{enumerate}
\item We designed a cognitive radio system with a cross-layer protocol that integrates DSA, backpressure algorithm and network coding.
\item We designed a distributed coordination mechanism with neighborhood discovery, spectrum sensing, channel rendezvous and local information exchange to support DSA, backpressure algorithm and network coding.
\item We integrated DSA and backpressure algorithm with network coding to support both unicast and multicast traffic by considering full rank and rank deficient decoding.
\item We designed a fully distributed system architecture on individual SDRs running their own versions of identical software.
\item We implemented the full protocol stack with USRP N210 radios using GNU Radio and Python modules.
\item We built a  repeatable and reconfigurable T\&E system based on a network channel emulator to assess the functionalities of the proposed cognitive radio system
\item We executed network emulation tests with USRP N210 radios and evaluated the performance under various network scenarios.
\end{enumerate}

The rest of the paper is organized as follows.
Sec.~\ref{sec:system} presents the system model.
Sec.~\ref{sec:distributed} describes the distributed coordination phases.
Sec.~\ref{sec:backpressure} presents the distributed backpressure algorithm for joint DSA and routing.
Sec.~\ref{sec:networkcoding} describes the integration of the backpressure algorithm with network coding.
Sec.~\ref{sec:radio} presents GNU Radio implementation and PHY layer aspects.
Sec.~\ref{sec:emulation} provides emulation test results with real radios.
Sec.~\ref{sec:con} concludes the paper.

\section{System Model}\label{sec:system}
A wireless multi-hop network of cognitive radio nodes is considered. Each node may act as the source, the destination or the relay for a packet flow. In unicast traffic, each node holds a separate queue for every packet flow identified by its source and its destination. In multicast traffic, each node holds a set of separate queues for every packet flow  that is identified by its source and the set of its destinations. There is a set of channels available for data transmission and control information exchange.

\begin{figure}
   \centering
   \includegraphics[width=1\columnwidth]{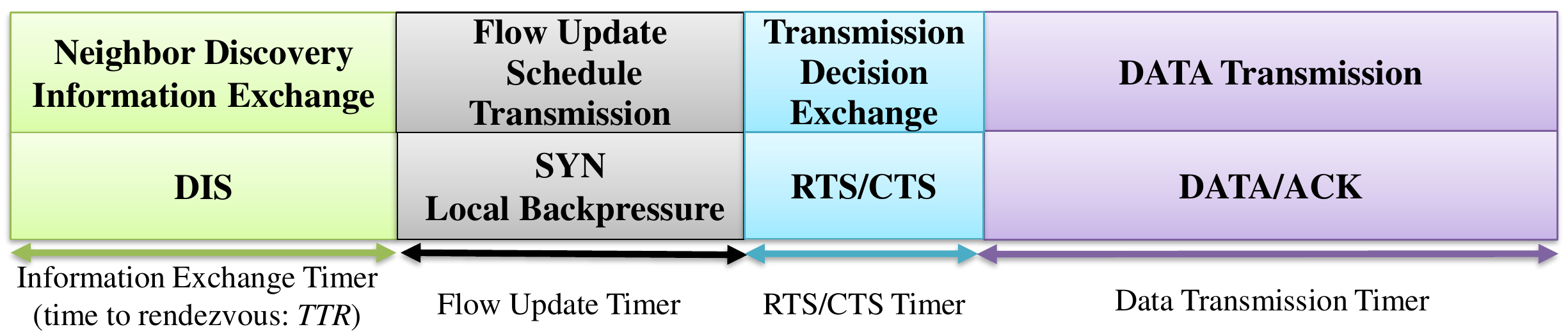}
   \caption{Four phases of distributed coordination.}\label{fig:2}
 \end{figure}

There is no common control channel or centralized controller. Instead, nodes share data channels for data or control packet transmissions, and thus adapt to network dynamics by discovering the local neighbor relations and updating the channel and queue information.
Each node individually applies \emph{distributed coordination} functions in four phases (see Fig.~\ref{fig:2}):
\begin{enumerate}
  \item neighborhood discovery and channel estimation;
  \item exchange of flow information updates and execution of the backpressure algorithm;
  \item transmission decision negotiation; and
  \item data transmission (including network coding).
\end{enumerate}

\begin{figure*}
   \centering
   \includegraphics[width=1.9\columnwidth]{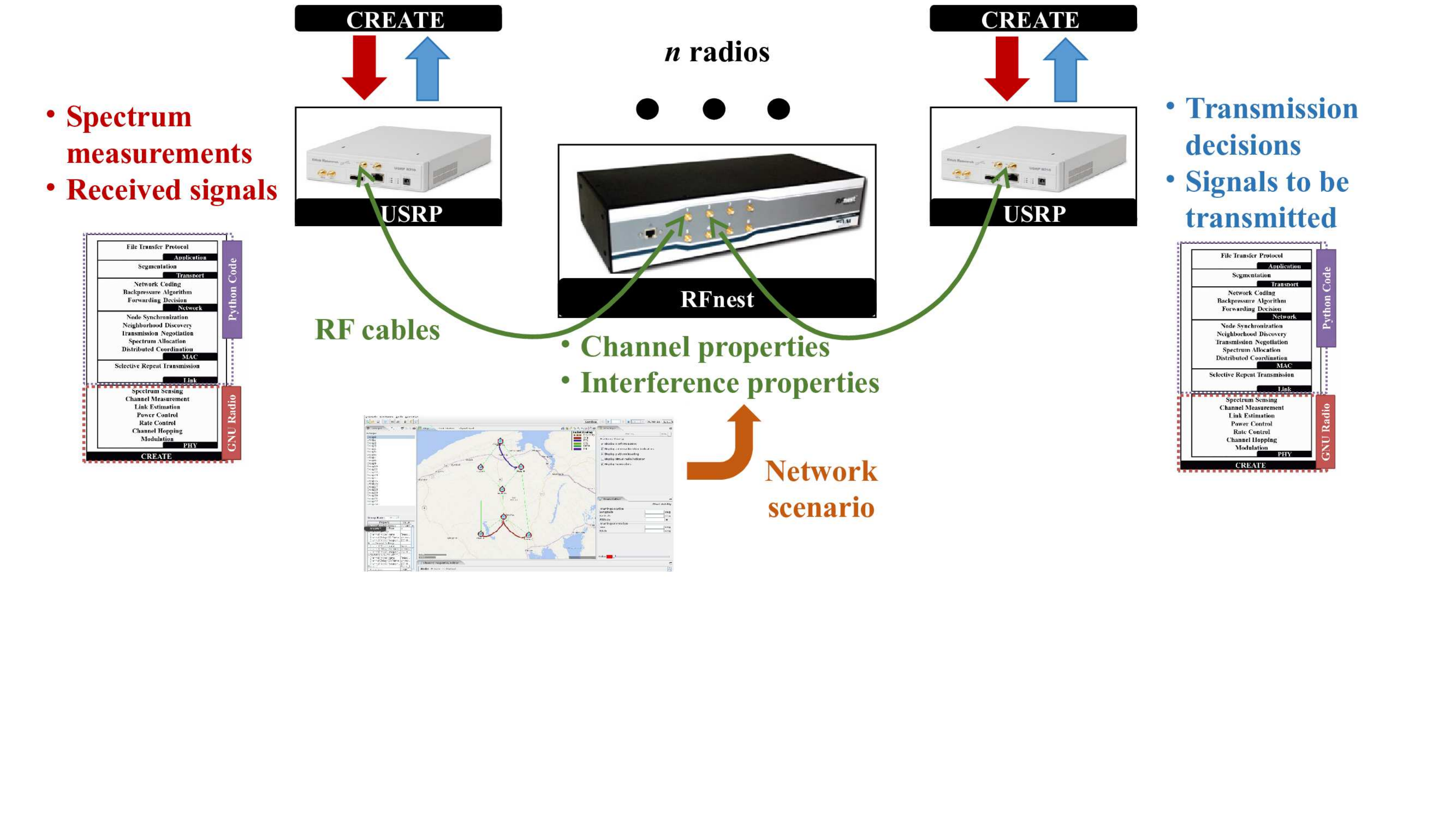}
   \vspace{-20mm}
   \caption{Cognitive radio system architecture.}\label{fig:system}
 \end{figure*}

In radio implementation, each node is represented by one SDR (USRP N210) and runs the same protocol stack (i.e., the identical radio code). Nodes communicate with each other over emulated channels using the network channel emulator, RFnest. Fig.~\ref{fig:system} shows the system architecture.

Each node aims to maximize its spectrum utility that combines link rate and traffic congestion. For that purpose, each node applies a combination of backpressure algorithm (for joint spectrum allocation and routing) and network coding. Fig.~\ref{fig:protocol} shows the protocol stack. Link/MAC, network, transport and application layers of the protocol stack are developed as Python modules, and Physical (PHY) layer is developed in GNU Radio. In addition to the application and transport layers (such as generating application traffic flows), Network, MAC and PHY layers implement the following capabilities:

 \begin{figure}
 \centering
 \includegraphics[width=0.75\columnwidth]{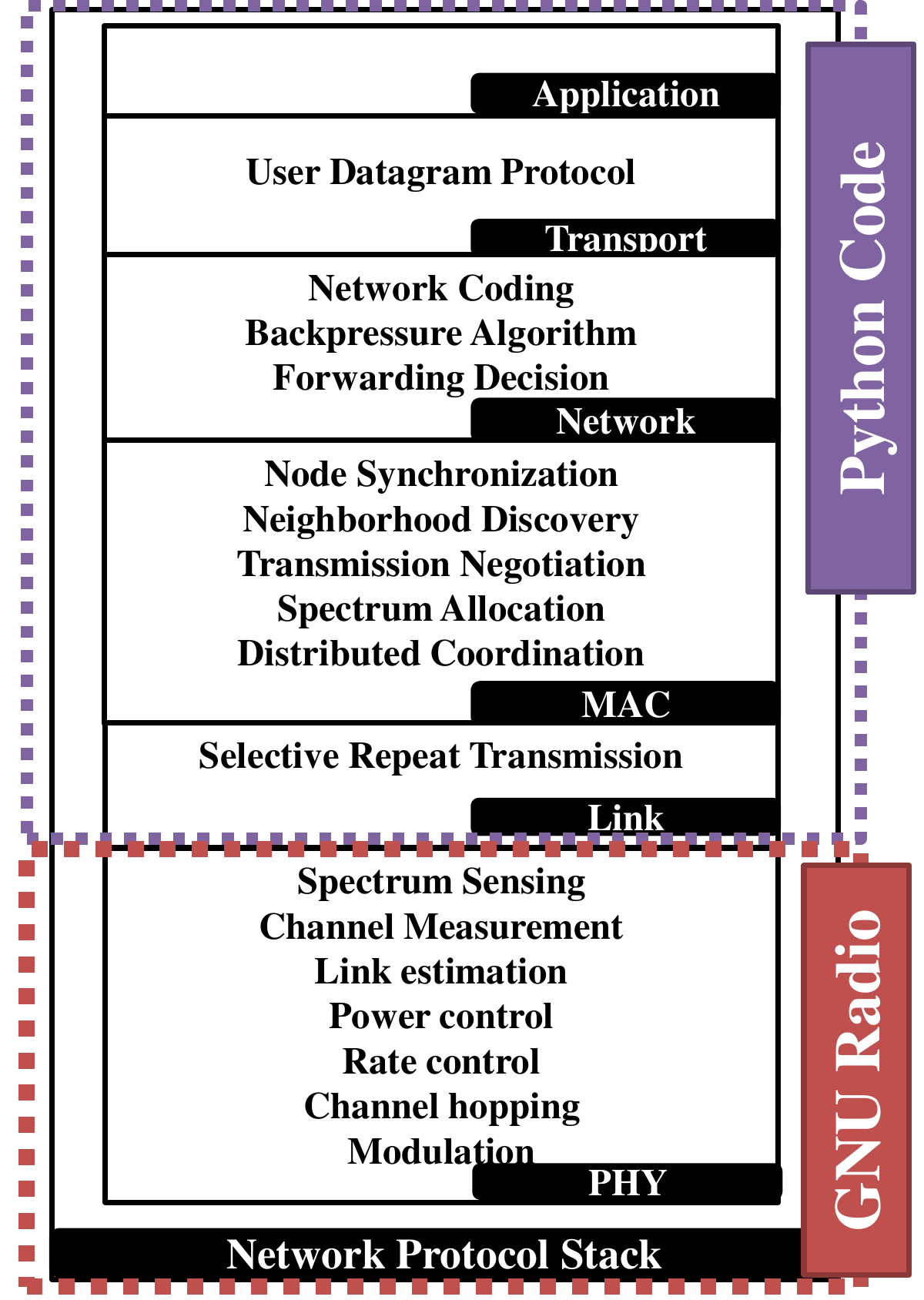}
 \caption{Cognitive radio network protocol stack.}
 \label{fig:protocol}
 \end{figure}

 \begin{figure}
   \centering
   \includegraphics[width=1\columnwidth]{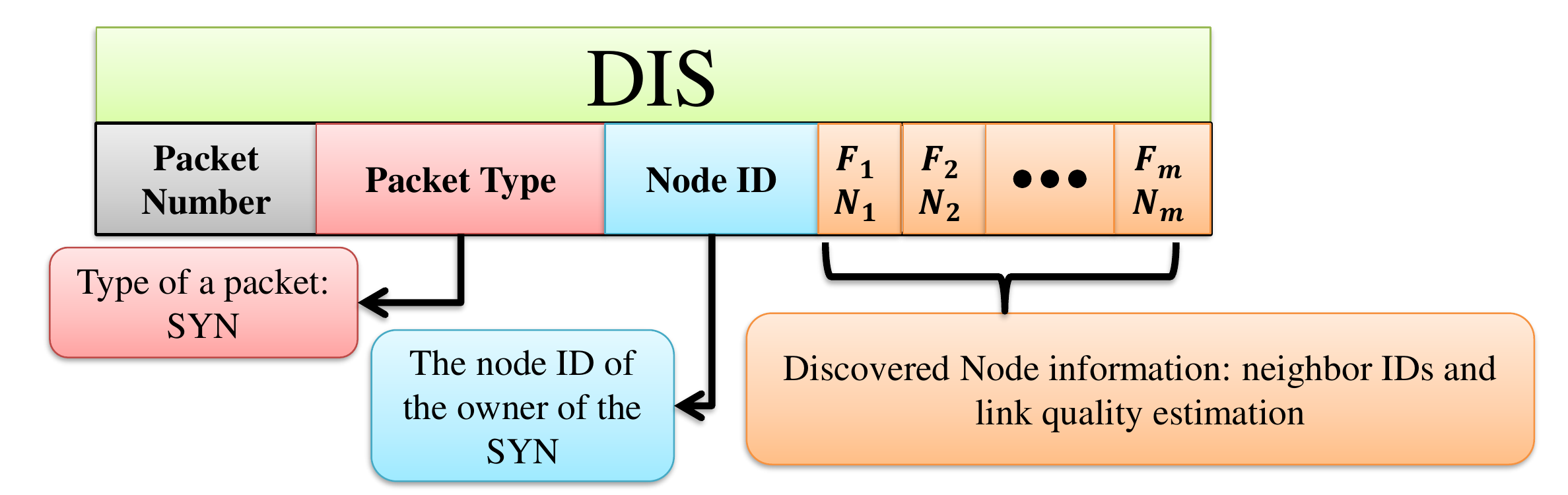}
   \caption{Format of discovery packet that is transmitted during the neighborhood discovery.}\label{fig:3}
 \end{figure}

\begin{enumerate}
\item \textit{Network layer}: Backpressure routing and network coding.
\item \textit{MAC layer}:
Neighborhood discovery, transmission negotiation, spectrum allocation, and distributed coordination functions.
\item \textit{PHY layer}: Spectrum sensing, channel measurement, link estimation, power control, channel hopping, modulation, and orthogonal frequency-division multiplexing (OFDM) transmission.
\end{enumerate}

\section{Distributed Coordination Phases of the Cognitive Radio System}
\label{sec:distributed}
A dedicated common control channel is a typical practice used in cognitive radio architectures  \cite{Lo11:control} to collect and share local information, and collectively compute strategies across the protocol stack. However, the common control channel poses a single point of failure and limits the rate of information exchange (due to its limited bandwidth). Instead, we consider the practical case in which nodes asynchronously establish distributed negotiation and coordination channels that are shared with data communications. Nodes use these channels to discover the local neighbor relations and update the channel and queue information. Each node individually goes though the following four phases.

\subsection{Phase 1: Neighborhood Discovery and Channel Estimation}
A node enters Phase 1 when it joins the network for the first time and then later periodically. By hopping among channels, the node visits a specific channel for a given period of time, broadcasts \emph{discovery} (DIS) messages to inform potential neighbors of its presence, and then switches to another channel. Fig.~\ref{fig:3} shows the DIS packet format that consists of:
\begin{enumerate}
\item the neighbor ID $i$ and
\item  the estimation of the channel quality for the link, which is specified by channel frequency $F_i$ in the next DIS packet.
\end{enumerate}
Each node estimates channels by calculating the root mean square (RMS) of the received signal power from the neighbor $i$'s DIS packet.

A new DIS packet is broadcast by a node when it switches to a new channel. This packet includes the information of those neighbors that this node has already discovered, and listens over the channel to receive DIS packets of new neighbors. Each node remains in the neighborhood discovery phase for a period of $\textit{TTR}$ (time-to-rendezvous), where $\textit{TTR} = F \times C$ seconds assuming each node visits $F$ channels, each for $C$ seconds. Whenever $\textit{TTR}$ expires, the node proceeds with the flow update and schedules the transmission phase. The channel sequence can be random, which is robust to changes in available channels and nodes joining/leaving the network. However, when the number of channels is large, a random hopping scheme may take long time to rendezvous. Thus, more sophisticated schemes, such as one based on Chinese remainder theorem \cite{Rahman}, can be applied to reduce the overhead. We consider only three channels in numerical results. Therefore, random channel hopping can be applied.
At the end of the neighborhood discovery phase, a \emph{local topology database} is built by each node and this database contains the list of the neighbors (that the nodes has discovered) and the channel quality of the links to those neighbors.

\subsection{Phase 2: Exchange of Flow Information Updates and Execution of Backpressure Algorithm}

\begin{figure}
   \centering
   \includegraphics[width=1\columnwidth]{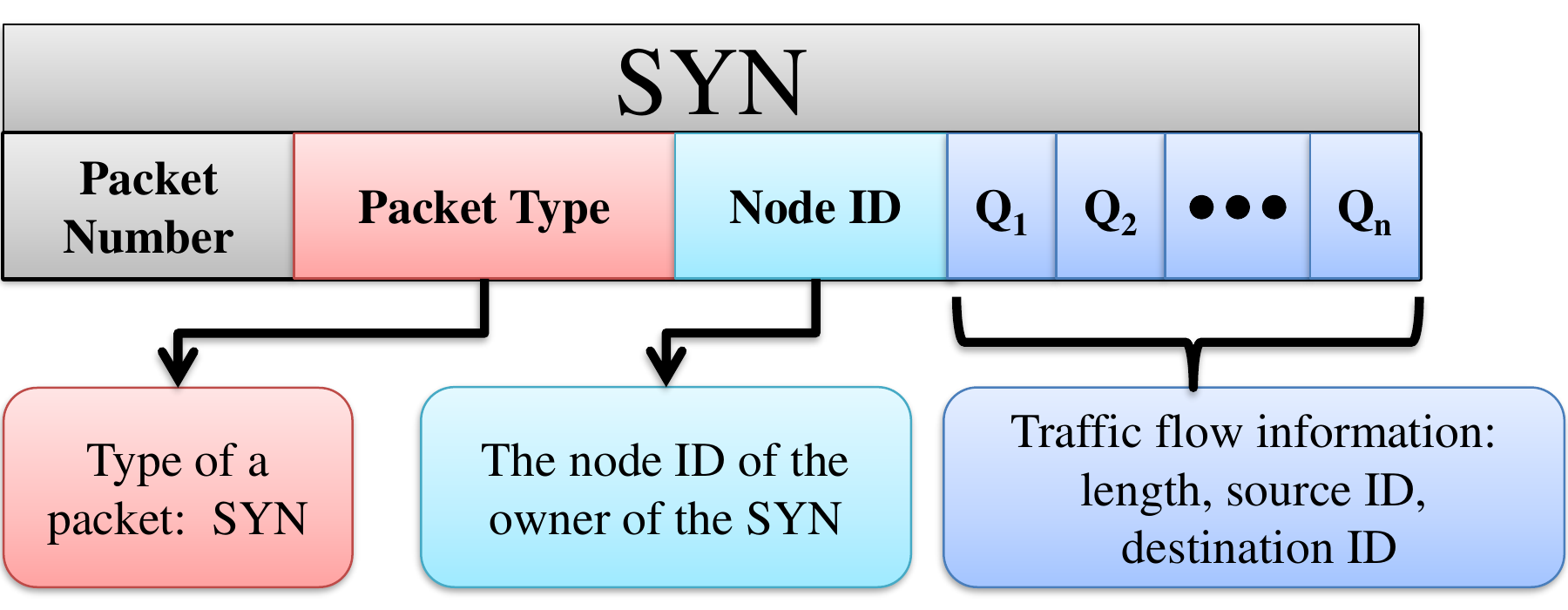}
   \caption{Format of SYN packet that is transmitted during the flow update.}\label{fig:4}
 \end{figure}

In Phase 2, a node hops among channels such that it visits a specific channel for a given period of time, broadcasts the SYN packet to inform its neighbors about the status of its queue backlog, and then switches to another channel. Fig.~\ref{fig:4} shows the SYN packet format that consists of:
\begin{enumerate}
\item the number of traffic flow queues; and
\item the number of packets in each queue (for a traffic flow identified by source and destination IDs).
\end{enumerate}

After receiving SYN packets from other nodes, each node iteratively builds a database, which includes
\begin{enumerate}
\item the queue information of the neighbors; and
\item the link quality of those channels, where the neighbors are available.
\end{enumerate}

If a node receives a SYN packet from its neighbor that is not included in its topology database (constructed during Phase 1), it estimates the channel quality and updates its topology database. After collecting queue length and channel information about its neighbors, each node notifies its network layer (through cross-layer feedback from MAC to network layer) to run the backpressure algorithm. The node proceeds with the transmission decision exchange phase if the backpressure algorithm finds the optimal transmission schedule. Otherwise, the node remains in the flow information update phase until it collects the necessary information for the backpressure algorithm. At the network layer, the backpressure algorithm
\begin{enumerate}
\item makes joint decisions of spectrum and power allocation;
\item selects the next hop for routing; and
\item coordinates transmissions for distributed channel access.
\end{enumerate}

We will present details of the backpressure algorithm in Sec. \ref{sec:backpressure} and its extension with network coding in Sec. \ref{sec:networkcoding}.

\subsection{Phase 3: Transmission Decision Negotiation}

 \begin{figure}
   \centering
   \includegraphics[width=1\columnwidth]{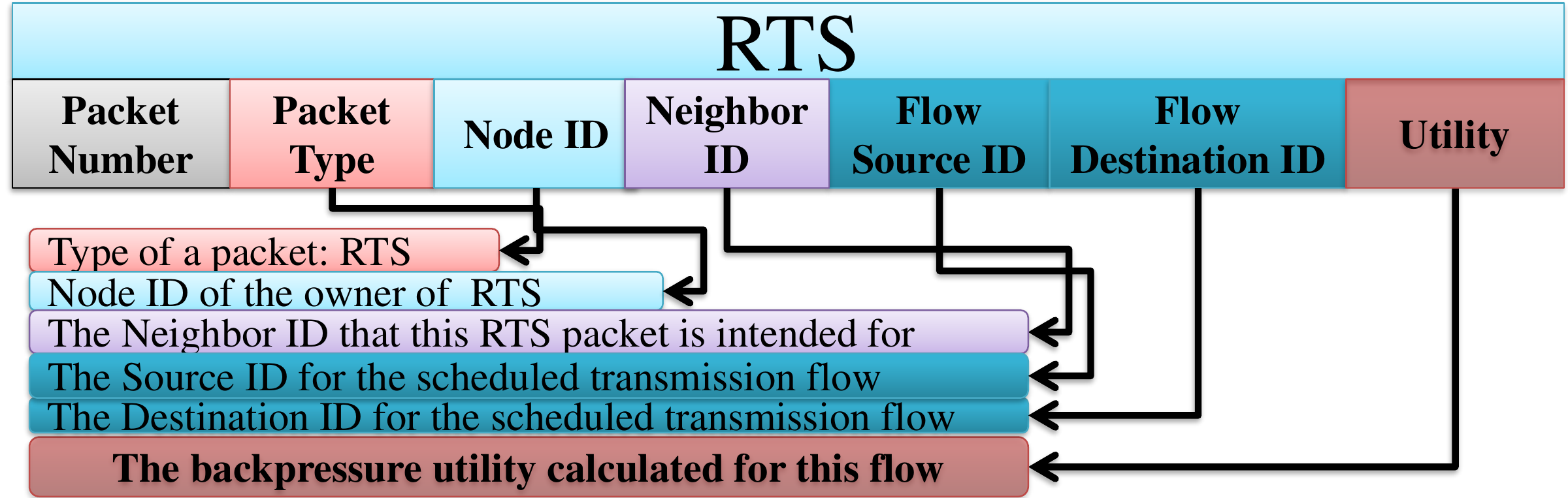}
   \caption{Format of RTS packet that is transmitted during the transmission decision exchange.}\label{fig:5}
 \end{figure}

\begin{figure}
   \centering
   \includegraphics[width=1\columnwidth]{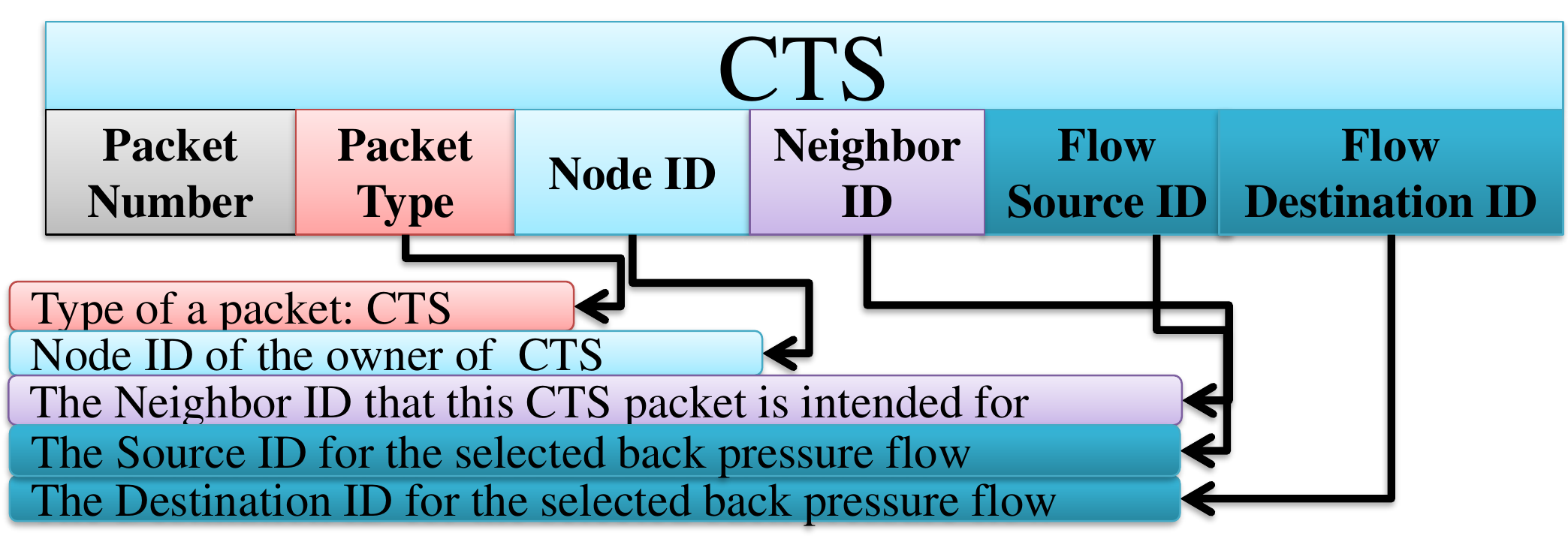}
   \caption{Format of CTS packet that is transmitted during the transmission decision exchange phase.}\label{fig:6}
 \end{figure}

In phase 3, each node transmits Request-To-Send (RTS) packet back to its neighbor over the link/channel determined by the backpressure algorithm.
Fig.~\ref{fig:5} illustrates the RTS packet format. Each node also listens over that channel for any possible RTS packets from its neighbors. After transmitting and potentially receiving RTS packets, the underlying decisions may conflict with each other.
We will solve the following three types of packet conflicts:
\begin{enumerate}
\item Two or more nodes have decided to transmit packets to each other at the same frequency channel. This corresponds to a half-duplex conflict assuming that a node cannot simultaneously transmit and receive packets.
\item Two or more nodes have decided to transmit packets to the same destination at the same frequency channel.
\item There exists a possible hidden node interference problem, i.e., a transmitter node is visible from the receiver node, but not from other nodes communicating with that receiver node.
\end{enumerate}
These conflicts are solved by considering the utility of each node's decision and selecting a node with higher spectrum utility (to be described in Sec.~\ref{sec:backpressure}) as the winner of the conflict. This way, some nodes  give up their decisions and let other nodes proceed with the transmission. Note that this conflict resolution does not need a centralized controller since each node has the complete information of its  decisions and its neighbors' decisions. Therefore, each node analyzes the received RTS packets and decides to receive data packets or insist on its own scheduled transmission. If it decides to receive packets, it transmits Clear-To-Send (CTS) packet to the neighbor that has the highest spectrum utility (to be described in Sec.~\ref{sec:backpressure}). Otherwise, it ignores the neighbor's RTS packet and does not send CTS packet back to that neighbor.
Fig.~\ref{fig:6} shows the CTS packet format.

A node stays in Phase~3 for a period of Transmission Decision Time ($\textit{TDT}$). If it receives a CTS packet in the meantime, it switches to the data transmission phase. Otherwise, it switches to the flow information update phase. If a node receives a CTS packet after it switches to the flow update phase, it switches back to the data transmission phase and initiates data transmission.
\subsection{Phase 4: Data Transmission}
\begin{figure}
   \centering
   \includegraphics[width=1\columnwidth]{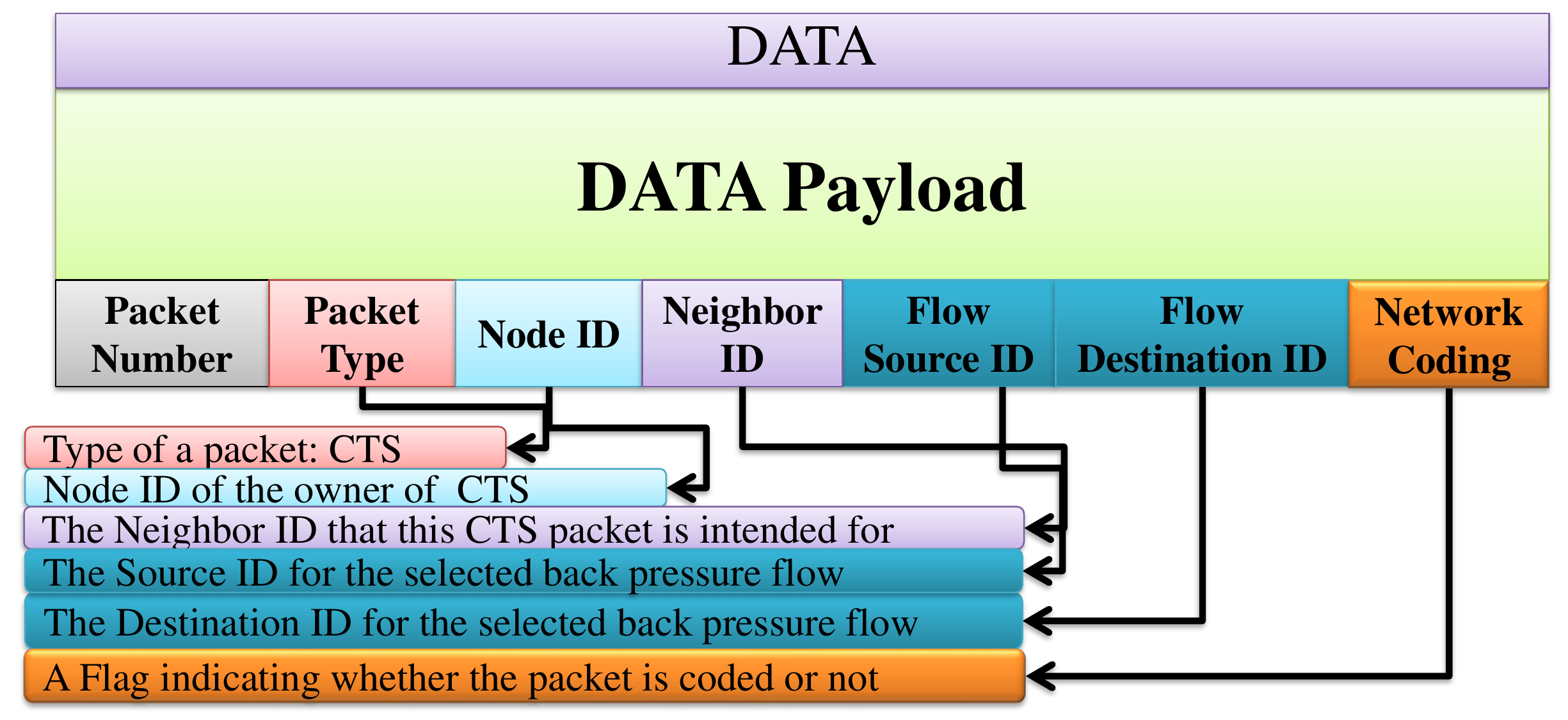}
   \caption{Format of DATA packet that is transmitted during the data transmission phase.}\label{fig:7}
 \end{figure}

A node enters Phase 4 to transmit or receive data packets. To transmit packets, a node gets data packets from the selected queue at the transport layer and generates a DATA packet. When network coding is applied, the node encodes packets before transmitting them. We present details of network coding in Sec. \ref{sec:networkcoding}. User Datagram Protocol (UDP) packet transmission is implemented at the transport layer. Fig.~\ref{fig:7} shows the DATA packet format.

\section{Backpressure Algorithm for Joint Spectrum Access and Routing} \label{sec:backpressure}
Each node $i$ keeps a separate queue for each flow $s$ and the backlog of that queue is $Q_i^s (t)$  at time $t$. For each link $(i,j)$, node $i$ chooses the flow to transmit as the one with the maximum difference of queue backlogs at the receiving and transmitting ends, namely
\begin{eqnarray}
s_{i,j}^*= \argmax_s  [Q_i^s (t)-Q_j^s (t)]^+, \label{eq:flow1}
\end{eqnarray}
where $[\cdot]^+ = \max⁡(\cdot,0)$.
Note that $Q_j^s (t)$ values are obtained from the SYN packets from node $j$. The spectrum utility for node $i$ transmitting to node $j$ is defined as
\begin{eqnarray}
U_{ij} (t)=c_{ij} (t)\left[Q_i^{s_{i,j}^*} (t)-Q_j^{s_{i,j}^*} (t)\right]^+, \label{eq:utility1}
\end{eqnarray}
 where $c_{ij}(t)$ is defined as the rate on link $(i,j)$ at time $t$ as an empirical throughput over link $(i,j)$. In particular, $c_{ij}(t)$ is computed from bit error rate, \textit{BER}, which follows from the empirical signal-to-noise-ratio, \textit{SNR}, obtained from the received signal power and the rate of the OFDM waveform based on the bandwidth and fast Fourier transform (FFT) symbol length. Next, node $i$ selects the next-hop neighbor and transmits to neighbor $j^* (t)$ that yields the maximum spectrum utility, i.e.,
\begin{eqnarray}
j^* (t) = \argmax_{j \in N_i}⁡ U_{ij}(t), \label{eq:link1}
\end{eqnarray}
where $N_i$ is the set of the next hop candidates for node $i$. 

The backpressure routing works well with heavy loads, namely when there is enough backlog to drive routing decisions. However, it may results in loops under light traffic. Therefore, we introduce a penalizing parameter $\alpha_j^s(t)$
to prevent a packet of flow $s$ from traversing the same node $j$ repeatedly. However, note that loops are not entirely inefficient and may be even necessary under heavy network traffic. The reason is that the underlying routing decisions are dynamic and it is possible that a path may not appear in previous iterations but it may become available later due to mobility.
The  parameter $\alpha_j^s(t)$ is initialized to $1$. Each time a flow $s$ revisits a node $j$ at time $t$, the parameter $\alpha_j^s(t)$ is updated by $1/f_j^s$ to discourage redundant paths, where $f_j^s(t)$ is the frequency that a flow $s$ has visited a particular neighbor $j$ up to time $t$.
Including $\alpha_j^s$, the flow from node $i$ to $j$ is selected as
\begin{eqnarray}
s_{i,j}^*= \argmax_s  \left\{ \left[Q_i^s(t) - Q_j^s(t)\right]^+ \alpha_j^s(t) \right\}  \label{eq:flow2}
\end{eqnarray}
that replaces (\ref{eq:flow1})
The link is scheduled according to (\ref{eq:link1}), where the spectrum utility in (\ref{eq:utility1}) is replaced by
\begin{eqnarray}
U_{ij} (t)=c_{ij} (t)\left[Q_i^{s_{i,j}^*} (t)-Q_j^{s_{i,j}^*} (t)\right]^+ \alpha_j^{s_{i,j}^*}(t) . \label{eq:utility2}
\end{eqnarray}

The outcome of the backpressure algorithm implemented at the network layer of each node $i$ is scheduling node $i$'s next transmission (if any) by specifying neighbor ID, channel frequency (link), and traffic flow.

\section{Extension with Network Coding} \label{sec:networkcoding}
This paper implements a practical network coding scheme operating on blocks of packets at the source and the intermediate nodes. Then the network-coded packets are decoded at the destination. \emph{Intra-flow} (intra-session) network coding is considered such that only packets of the same flow are coded with each other.
We make network coding practical with the following attributes.
\begin{enumerate}
\item \textit{RLNC}: Coding coefficients are uniformly randomly selected for distributed implementation without any centralized controller.
\item \textit{Packet tagging}: Each packet is tagged with the corresponding coding vector to support distributed implementation.
\item \textit{Buffering}: Asynchronous packet arrivals and departures are supported with buffering for  arbitrarily varying rate, delay, and loss.

\item \textit{Virtual multicast queues}: To support multicast traffic, each node implements one virtual queue
for each destination in the multicast group.
\end{enumerate}

We extend the backpressure algorithm to support \emph{multicast} data flows in two ways:
\begin{enumerate}
  \item  Virtual queues are used to store data for each multicast destination. Then multicast flows are supported by applying the backpressure algorithm  on these virtual queues.
  \item RLNC is combined with the backpressure algorithm and network-coded packets are decoded by two different methods, namely full rank and rank deficient decoding.
\end{enumerate}

\subsection{Packet Encoding}

Data (byte) symbols are grouped into packets at the source node (using their ASCII representation), i.e., the $i$-th packet is $\mathbf{x}_i=[x_{i,1},...,x_{i,N}]$. Following practical implementation of network coding (e.g., \cite{Chou}), network coding is applied at a source node or an intermediate node applies such that incoming packets are coded for its destination $d$ as
\begin{eqnarray}
\left[\mathbf{y}_1^{\text{T}} \cdots \mathbf{y}_h^{\text{T}} \right]^{\text{T}} = G_d \left[\mathbf{x}_1^{\text{T}} \cdots \mathbf{x}_h^{\text{T}} \right]^{\text{T}}.  \label{eq:yT}
\end{eqnarray}
The coefficients of $G_d$ are uniformly randomly selected. In (\ref{eq:yT}),  $(\cdot)^{\text{T}}$ is the transpose of a matrix or a row vector, the encoding matrix $G_d$ codes $h$ packets $\mathbf{x}_1 \cdots \mathbf{x}_h$ to obtain packets $\mathbf{y}_1 \cdots \mathbf{y}_h$.
When a node finds a transmission opportunity on one of its outgoing links, it transmits a coded packet.
At each intermediate node, the network coding operation (\ref{eq:yT}) is applied to the incoming packets and network coding matrix $G_d$ is updated and carried with data packets. Source packets are recovered by destination $d$ that inverts $G_d$ and applies the inverse to received packets. RLNC does not require any node to know the global network topology \cite{Ho1}. Only packets of the same flow are coded with each other.

We use Galois field arithmetic for encoding and decoding functions. While each packet contains up to 500 bytes, a block may be incomplete block. In that case, the block is filled before coding by using the ISO/IEC 7816-4 padding.
The remaining portion of the block is filled by padding the next byte with a hexadecimal x80 byte that falls outside the range of ASCII and is used as a signal byte. Then the block is completed by zero padding. Using only zero padding has the disadvantage that the data may end with zero values and the padding may be irreversible. Therefore, the end of the actual data is declared with a signal bit.

An encoding vector $\mathbf{g}_i$ (where $\mathbf{g}_i^{\text{T}} = G_d \: \mathbf{u}_i^{\text{T}})$ is used to tag every coded packet such that
\begin{eqnarray}
\left[ [\mathbf{g}_1,\mathbf{y}_1]^{\text{T}} \cdots [\mathbf{g}_h,\mathbf{y}_h]^{\text{T}} \right]^{\text{T}} = G_d \left[ [\mathbf{u}_1,\mathbf{x}_1]^{\text{T}} \cdots [\mathbf{u}_h,\mathbf{x}_h]^{\text{T}} \right]^{\text{T}}.
\end{eqnarray}

Packet tagging has the advantage that the destination(s) can find the encoding vectors necessary to decode the received packets within those packets. In the meantime, there is no need that the destination knows the encoding functions or the network topology, to compute $G_d$.

A continuous stream of packets is transmitted for each information source as blocks and $h$ source packets are used per block. In that case, all the coded packets related to the $k$-th block of source packets $\mathbf{x}_{kh+1},...,\mathbf{x}_{kh+h}$ belong to the generation $k$.
We keep track of packets in the same generation by tagging each packet with its \emph{generation number} $k$.
Then packets within a generation are synchronized in coding (at the source) and decoding (at the destination). Note that we consider digital network coding at the packet level compared with analog (or physical layer) network coding that relies on superposition of received signals \cite{Sagduyu7, Sagduyu11, Honig3}.

\subsection{Packet Decoding}
For decoding network-coded packets, a destination collects $h$ or more packets in a given generation, stacks their symbols row-by-row, extracts the symbols in the packet tags to form $G_d$, and applies the inverse of $G_d$, if it exists, to the symbols in the packet payloads. Equivalently, the destination can apply \emph{Gaussian elimination} on the matrix of symbols formed by the stacked packet tags and payloads.
The algorithmic decoding delay is the length of time for the destination to collect $h$ packets, which is proportional to the generation size $h$.

For low algorithmic decoding delay, we use \emph{earliest decoding} in Python and wrap it by GNU Radio. After receiving each packet, \emph{Gaussian elimination (from full rank matrix)} is applied to the matrix of symbols that are formed by the packets stacked so far. As $G_d$ tends to be lower triangular (i.e., the $i$-th received packet tends to be a linear combination of the first $i$ source packets because of the causality of computation in the network), it is typically possible to decode the first $i$ source packets after receiving more than $i$ coded packets.

Alternatively, we also used \emph{rank deficient decoding} \cite{Yan}. A full rank decoder requires a threshold of linearly independent equations accumulated before decoding can begin. On the other hand, rank deficient decoding can start as soon as row vectors that are not necessarily dependent are received. Rank deficient decoding improves the delay of decoding and combined with RLNC, it improves reliability and throughput \cite{Yan}. By using the column vectors of the received data, rank deficient decoding determines values of a particular row vector in the same column position.
The packets are encoded linearly through matrix multiplication $\mathbf{y} = G_d \mathbf{x}$, where the received encoded packets are represented by the rows of $\mathbf{y}$,  $G_d$ (known at the destination) represents the coefficient matrix for encoding and $\mathbf{x}$ is the data payload. As packets keep arriving, the destination computes the inverse by using the matrix $G_d$. Consider a particular column position $l$. Then the columns $ W_l$ of the decoded matrix $W$ is recovered from a linear system of equations in the form of $V_l  = G_d W_l$. This operation is parallelized across all columns. Then all of the columns along a row vector are combined to recover each data packet.

\subsection{Integration of Network Coding with Backpressure Algorithm}
Network coding is modified for use with the backpressure routing as follows. We consider intra-flow network coding with  \emph{virtual queues} \cite{Ho2}. In this setting, each node maintains the physical packet queue as well as virtual queues for destinations in multicast group.

Packets are coded in each virtual queue independently with different random matrices. This approach supports better recovery from losses by offering more diversity to the same data.
Utility computation of backpressure algorithm uses the backlog of each virtual queue to schedule packet transmissions supported by decision negotiation process through RTS and CTS message exchanges. In the special case of unicast traffic, the utility is summed over only one virtual multicast queue.

Virtual queues are used to extend the backpressure algorithm for the multicast traffic. Node $i$ selects the flow to $j$ from
\begin{eqnarray}
s_{i,j}^*= \argmax_s  \left\{ \sum_{d} \left[Q_i^{s,d}(t) - Q_j^{s,d}(t)\right]^+  \alpha_i(t) \right\}, \label{eq:sij}
\end{eqnarray}
where $Q_i^{s,k}(t)$ is the length of virtual queue at node $i$ maintained for destination node $k$ in multicast session $d$.
In (\ref{eq:sij}), backpressure algorithm is extended to multicast flows with virtual queues by summing the differential backlogs over all destinations. After node $i$ selects next hop neighbor $j$ and the flow $s_{i,j}^*$ to transmit, $i$ and $j$ exchange RTS and CTS to finalize the transmission decision. The packet is removed from the virtual queue for all destinations for which the next-hop neighbor is selected and only one packet is transmitted. A virtual queue is removed once the corresponding destination is reached.
A destination may be a relay node. In that case, virtual queues for other destinations are still kept.

In the data transmission phase to transmit or receive data packets, a node gets its data packets from the corresponding queue at the transport layer and generates a data packet. Then the node transmits  network-coded packets. UDP packet transmission is implemented at the transport layer.

\section{SDR Implementation}
\label{sec:radio}

\subsection{SDR Modules}

\begin{figure*}
   \centering
   \includegraphics[width=1.75\columnwidth]{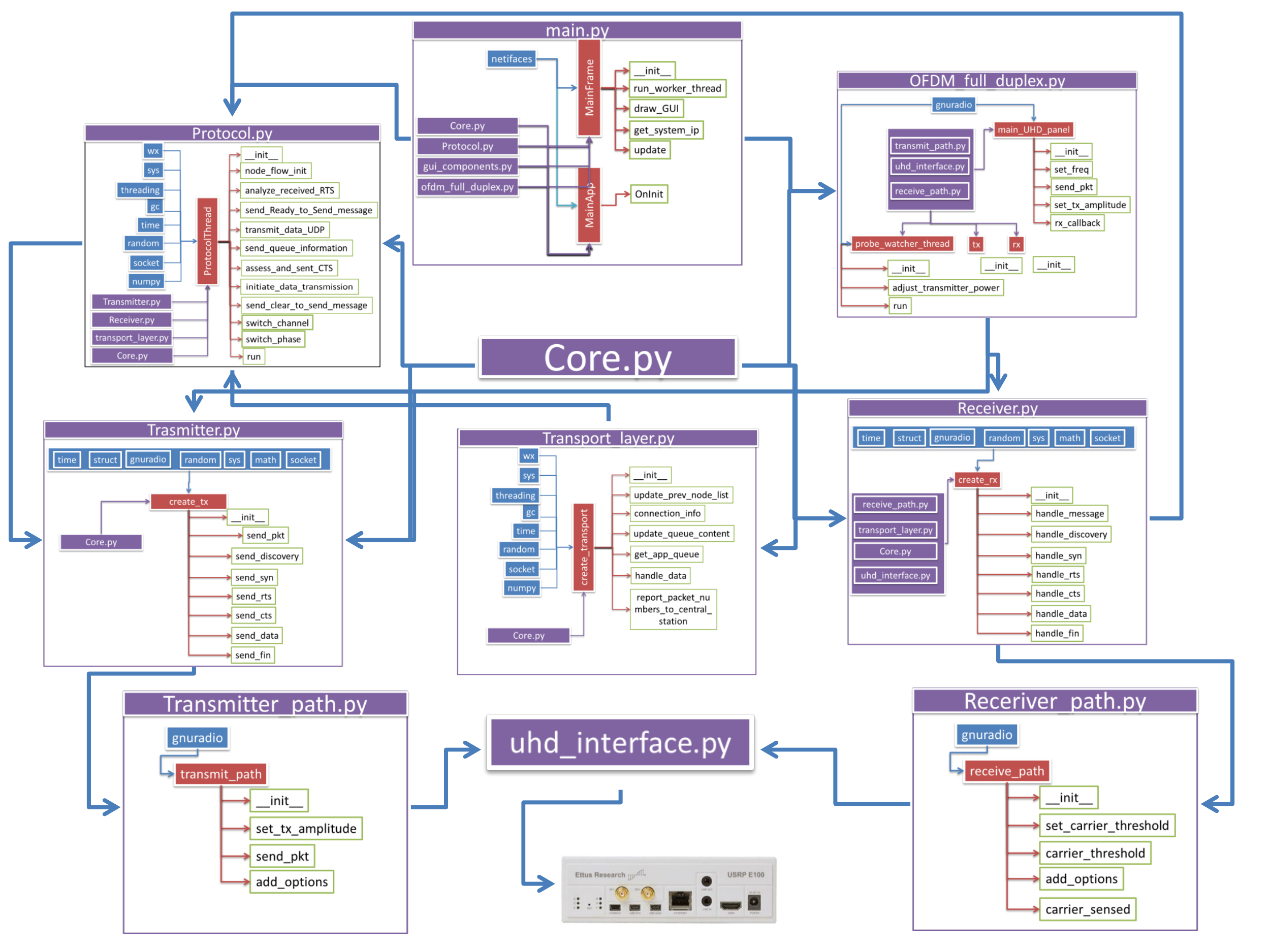}
   \caption{GNU Radio code framework.}\label{fig:framework}
 \end{figure*}

Fig.~\ref{fig:framework} shows the code framework for the SDR implementation.
The \texttt{main.py} module  provides capabilities on system initialization and system update, which is based on \texttt{Core.py}, \texttt{Protocol.py} and \texttt{OFDM\_full\_duplex.py} modules.

\begin{itemize}
  \item The \texttt{Core.py} module contains global variables, classes and data structures used by different Python files.
  \item The \texttt{Protocol.py} module implements various protocols, e.g., system, node, and flow initialization, analyze RTS, transmit CTS, transmit data, exchange queue information,  switch, channel and switch distributed coordination phase. This module is based on  \texttt{Core.py} module, \texttt{Transmitter.py} and \texttt{Receiver.py} modules, and \texttt{Transport\_layer.py} module.
      \begin{itemize}
        \item The \texttt{Transmitter.py} module provides transmitter capabilities, e.g., initialization, transmit discovery packet, transmit SYN packet, transmit RTS/CTS, and transmit data.
        \item The \texttt{Receiver.py} module provides receiver capabilities, e.g., initialization, handle message, handle discovery packet, handle synchronization packet, handle RTS/CTS, and handle data.
      \end{itemize}
  \item The  \texttt{OFDM\_full\_duplex.py} module implements the full duplex capability in OFDM, e.g., initialization, set frequency/amplitude, adjust transmit power, transmit packet, and receiver callback. This module is based on three modules: \texttt{Transmit\_path.py} module, \texttt{UHD interface.py} module, and \texttt{Receive\_path.py} module. Sec. \ref{subsec:phy} provide details of these modules.
\end{itemize}

\subsection{PHY Layer Implementation}
\label{subsec:phy}

 USRP N210 radios are used for implementation. They are equipped with SBX transceivers. For the PHY layer, we implemented the transmit and receive paths using GNU Radio. The transmit path is used to transmit the waveform and the receive path is used to decode packets from received waveform. Different communication options are possible, including modulation (BPSK, QPSK, etc.), OFDM symbol length (e.g., 512), OFDM occupied carrier length (e.g., 200), transmitter power (e.g., ranging from 0 to 31.5 dB), transmitter amplitude (e.g., scales from 0 to 1.0 relative to the maximum value), frequency carrier (e.g., SBX transceivers support 4.4MHz to 4.4 GHz), and receiver gain (e.g., ranging from 0 to 31.5 dB). While some parameters are set fixed, some of them (e.g., transmitter amplitude and frequency carrier) change over time along with protocol decisions. Below we describe transmit path, receiver path, real-time spectrum sensing, and power control components.

\begin{itemize}
\item \textit{Transmit Path}: Modulation parameters are extracted and the modulator class is initialized to modulate the data and deliver it to the \texttt{USRP} module for transmission. Then the \texttt{USRP} module is initiated with the following parameters: 1) \textit{Symbol rate} based on the transmitter bit rate and the modulation scheme. 2) \textit{Sample per symbol}, the number of bits per symbol. 3) \textit{Transmitter frequency}, the carrier frequency (channel) where the USRP will transmit. An instance of the transmit path generating the RF signal is connected to the USRP.

\item \textit{Receive Path}: The demodulator is built based on the user modulation choice and the modulation parameters. Then a filter is designed to get the actual channel to operate on. A low-pass filter is designed and the filter coefficients such as gain, the sampling rate, the center of the transmission band, the width of the transmission band and the filter type are specified. This is followed by defining a packet receiver to demodulate the received signal. The incoming signal is detected by the \texttt{Receive\_path}  module using carrier sensing block. The carrier sense block and the packet receiver are connected with the channel filter. Following the demodulation, the payload is delivered by the \texttt{Receive\_path}  module to higher layers.

    \item \textit{Real-Time Spectrum Sensing}:  Nodes are enforced by spectrum sensing to back off. The intention of each node is to transmit over an idle channel or to switch to another channel if the channel is busy. This way, the likelihood of signal interference and packet distortion are reduced significantly. Real-time spectrum sensing is implemented by using multithreading to sense the spectrum in real time (when the radio is not transmitting). The energy of the signal (calculated using the average magnitude square of the sensed signal) is returned.

\item \textit{Power Control}: Initially, each node $i$ transmits its packets with power $P_i (t)$ to meet the target SNR,  $\gamma _T$. Node $i$ extracts the estimation of the achieved SNR $\hat{\gamma}_i (t)$ from the SYN packets received from other neighbors. The transmission power of node $i$ is updated as $P_i (t+1)= P_i (t) \times \gamma_T \times (\hat{\gamma}_i (t))^{-1}$ for transmission at time $t+1$.
\end{itemize}

 \begin{figure*}
   \centering
   \includegraphics[width=1.75\columnwidth]{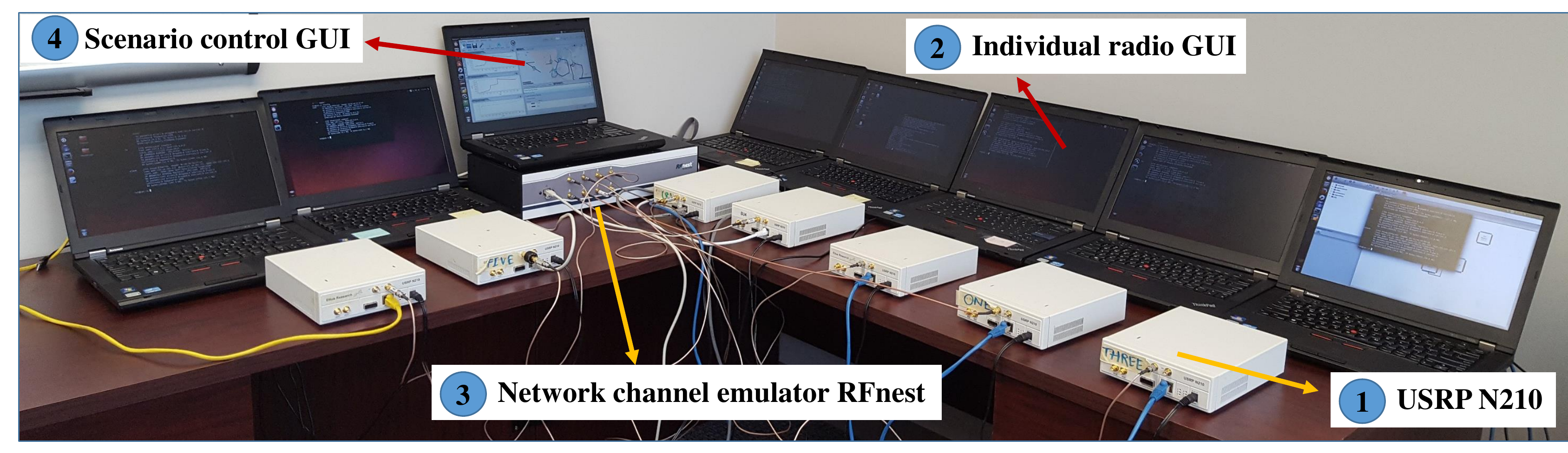}
   \caption{Cognitive radio network emulation testbed.}\label{fig:testbed}
 \end{figure*}

\section{Cognitive Radio System Evaluation}
\label{sec:emulation}

\subsection{T\&E System }

T\&E system is illustrated in Fig.~\ref{fig:testbed}. This system consists of seven USRP N210s \cite{Ettus}. The size of the network can scale up seamlessly with more radios. The front-end of each cognitive radio node is represented by one USRP radio. We implemented the PHY layer with GNU Radio \cite{Blossom} at each controller laptop, where link/MAC, network, transport and application layers are implemented by Python and integrated within the GNU Radio software. This T\&E system is plug-and-play and allows other radios to be integrated as primary users or as RF interferers. Different topologies with wireless mesh connection capability are supported by RFnest, a Field Programmable Gate Array (FPGA) based network channel emulator that emulates network channels in real time such that all nodes experience real channel impulse responses \cite{Yackoski}. In this emulation environment, we remove antennas of USRPs and connect them with RF cables to RFnest such that RF signals travel through RFnest. This way, we repeat all radio experiments with identical channel scenarios. Table \ref{table:1} provides the emulation test parameters.

 \begin{table}
 \caption{Radio parameters for network emulation tests.}
 \small
 \begin{center}
 \begin{tabular}{ll}
 \hline\noalign{\smallskip}
   \textbf{Parameter} & \textbf{Value} \\
 \noalign{\smallskip}\hline\noalign{\smallskip}
   Duration to visit a single channel	& 2 seconds \\
 Duration in discovery phase &	60 seconds \\
 Duration in decision negotiation phase &	60 seconds \\
 Duration to exchange data packets &	30 seconds \\
 Duration to broadcast queue updates	& 20 seconds \\
 Channels to visit	& 2.41, 2.43, 2.46GHz \\
 Packet size	& 500 bytes \\
 OFDM FFT length	& 512 \\
 OFDM cyclic prefix length &	128 \\
 OFDM occupied length &	200 \\
 OFDM modulation	& BPSK \\
 USRP power transmission range & (-15,-5) dBm \\
 \noalign{\smallskip}\hline
 \end{tabular}
 \end{center}
 \label{table:1}
 \end{table}

 \begin{figure}
   \centering
   \includegraphics[width=1\columnwidth]{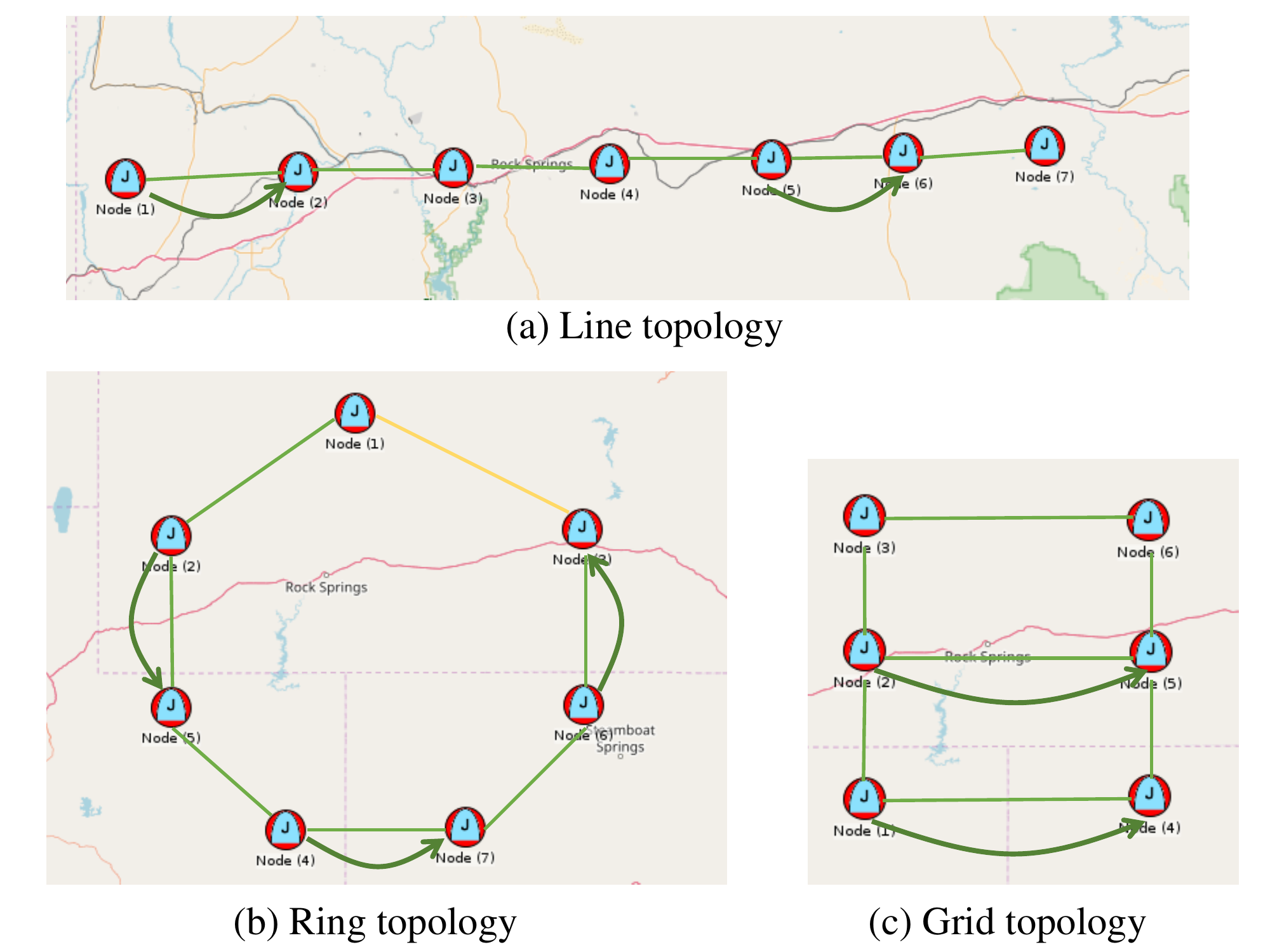}
   \caption{Topologies for emulation tests without network coding.}\label{fig:topology}
 \end{figure}

 \begin{figure*}
 	\centering
 	\subfigure[Average overhead.]
 	{\includegraphics[width=0.5\columnwidth]{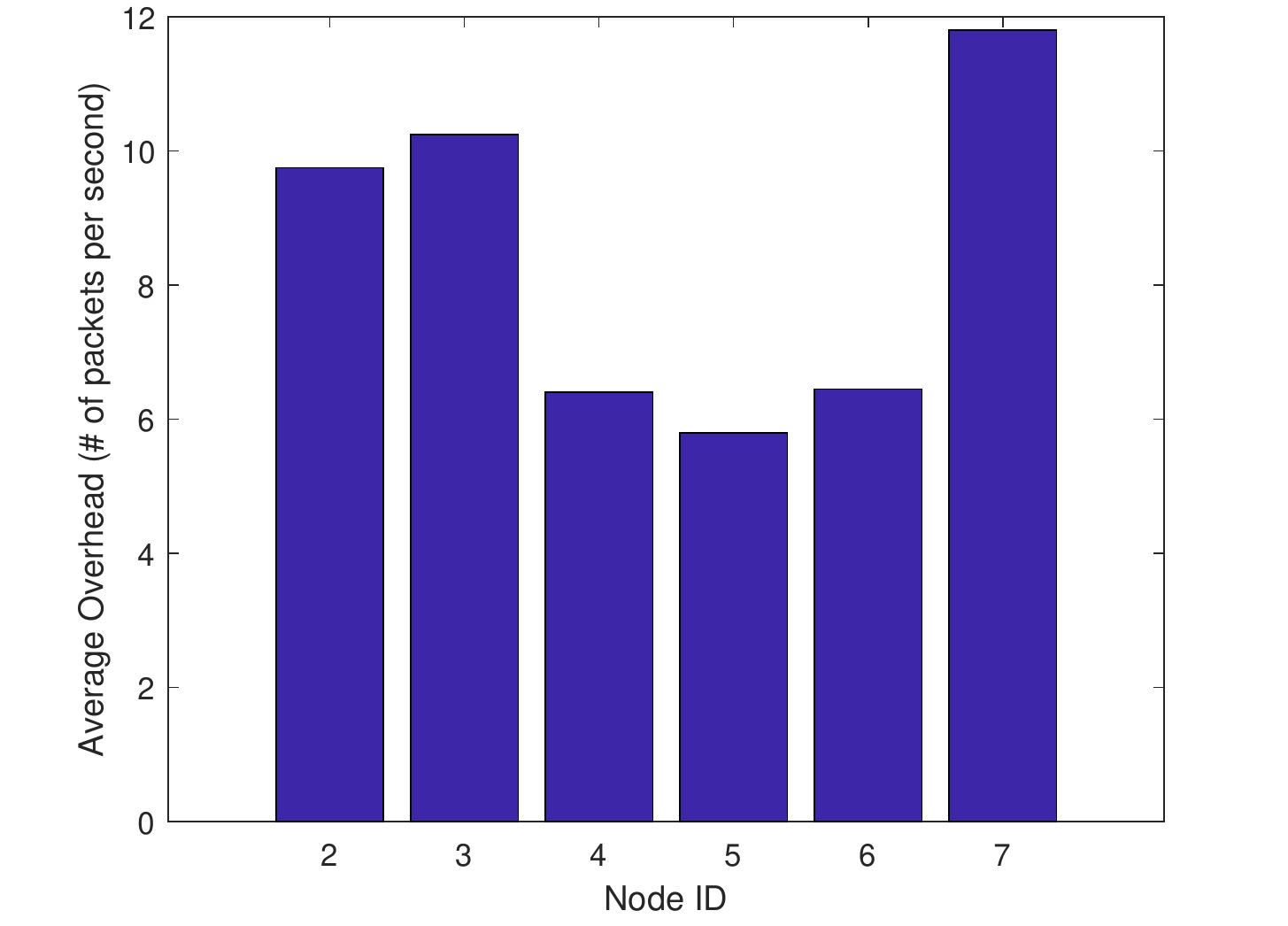}}
 	\hfil
 	\subfigure[Average energy consumption.]
 	{\includegraphics[width=0.5\columnwidth]{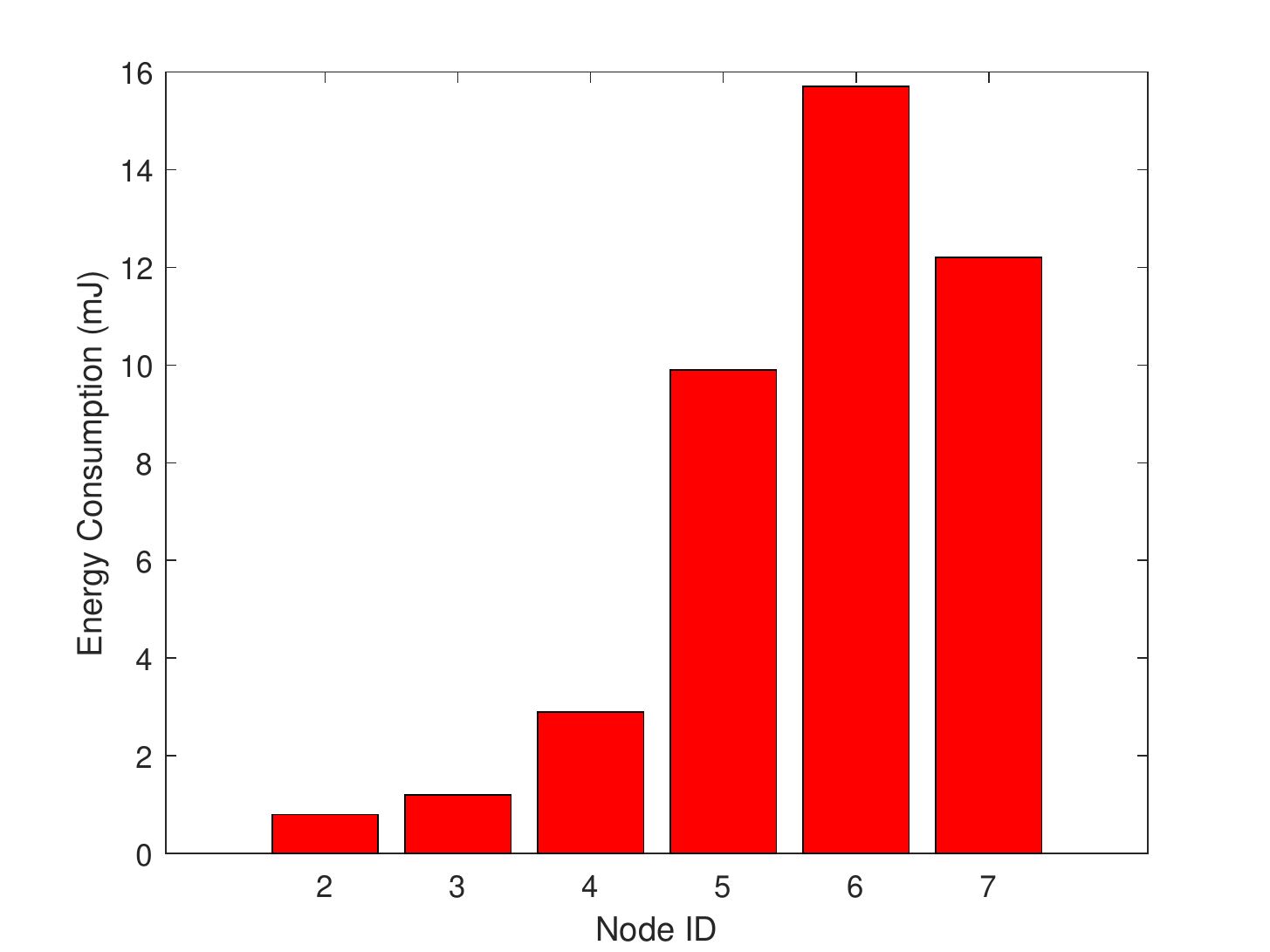}}
 	\hfil
 	\subfigure[Average backlog or throughput.]
 	{\includegraphics[width=0.5\columnwidth]{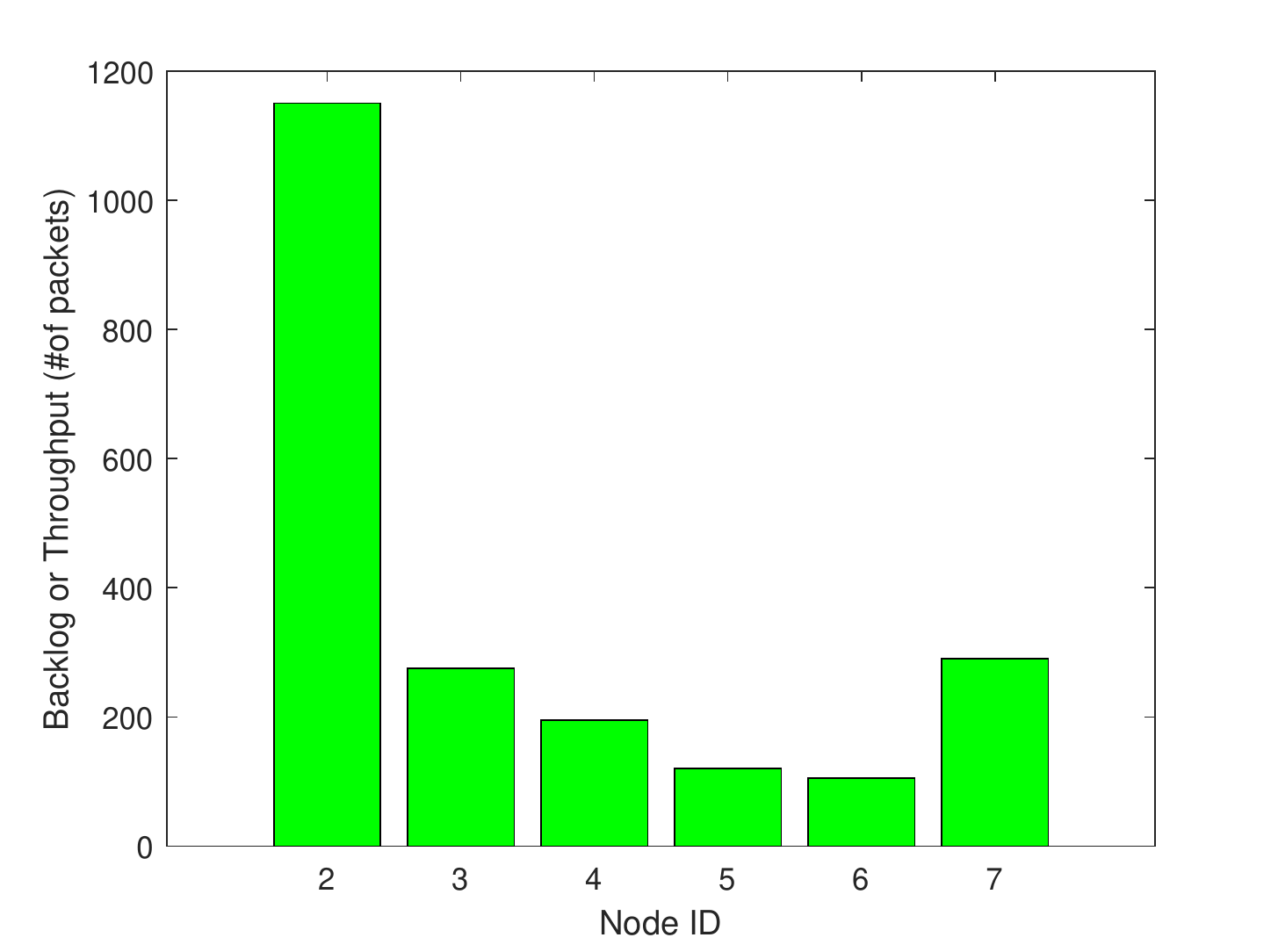}}
 	\caption{Performance for line topology.}
 	\label{fig:line}
 \end{figure*}

\subsection{Emulation Testbed Results without Network Coding}

Three scenarios (``line", ``ring" and ``grid") shown in Fig.~\ref{fig:topology} are considered, where strong links are colored as green and weak links are colored as yellow. We also use arcs to show concurrent transmissions in a channel.
We start with the case without network coding for unicast traffic. Node 1 is the source. Node 7 is the destination for line and ring topologies, and Node 6 is the destination for grid topology. Other nodes act as relays for others. We consider a single flow to simplify our discussion on performance and overhead. The testbed can support multiple flows.

The performance results for line topology are shown in Fig.~\ref{fig:line}. From Fig.~\ref{fig:line}-(a), we observe that Nodes 2, 3 and 7 incur higher overhead. This is expected since Nodes 2 and 3 are the first two relays after the source Node 1 and therefore should establish communication continuously with the source to receive data packets, while Node 7 as a destination transmits control packets periodically to receive its data packets. Therefore, these nodes incur higher overhead than the other relay nodes. From Fig.~\ref{fig:line}-(b), we observe that Nodes 5-7 have higher energy consumption than the other nodes. Fig.~\ref{fig:line}-(c) shows the average backlog for relay nodes (Nodes 2-6) and the throughput for destination (Node 7).  Node 2 has the higher backlog with respect to the other relay nodes, since it is the first relay node for the source (Node 1).

\begin{figure*}
     \centering
     \subfigure[Average overhead.]
     {\includegraphics[width=0.5\columnwidth]{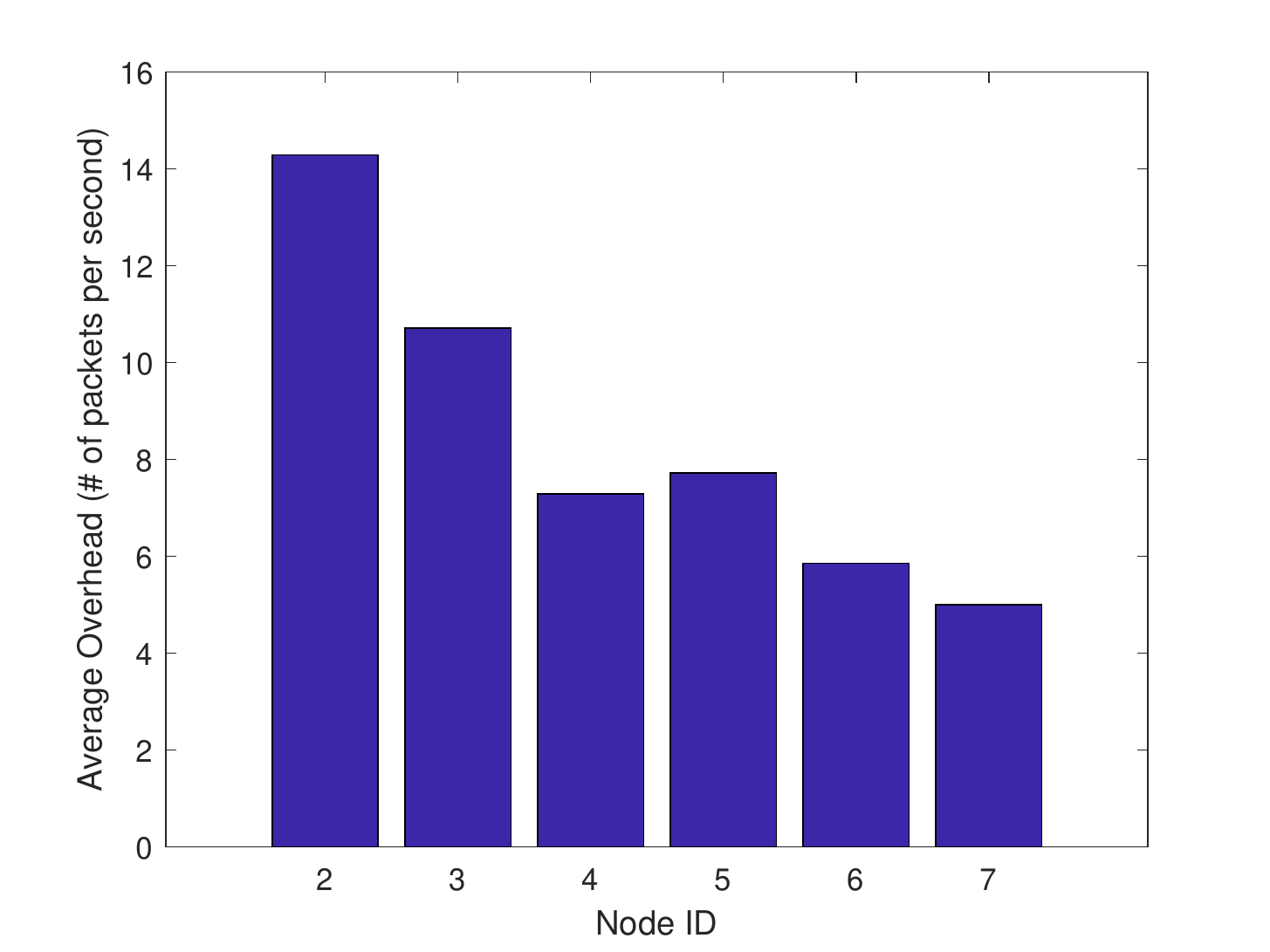}}
     \hfil
     \subfigure[Average energy consumption.]
     {\includegraphics[width=0.5\columnwidth]{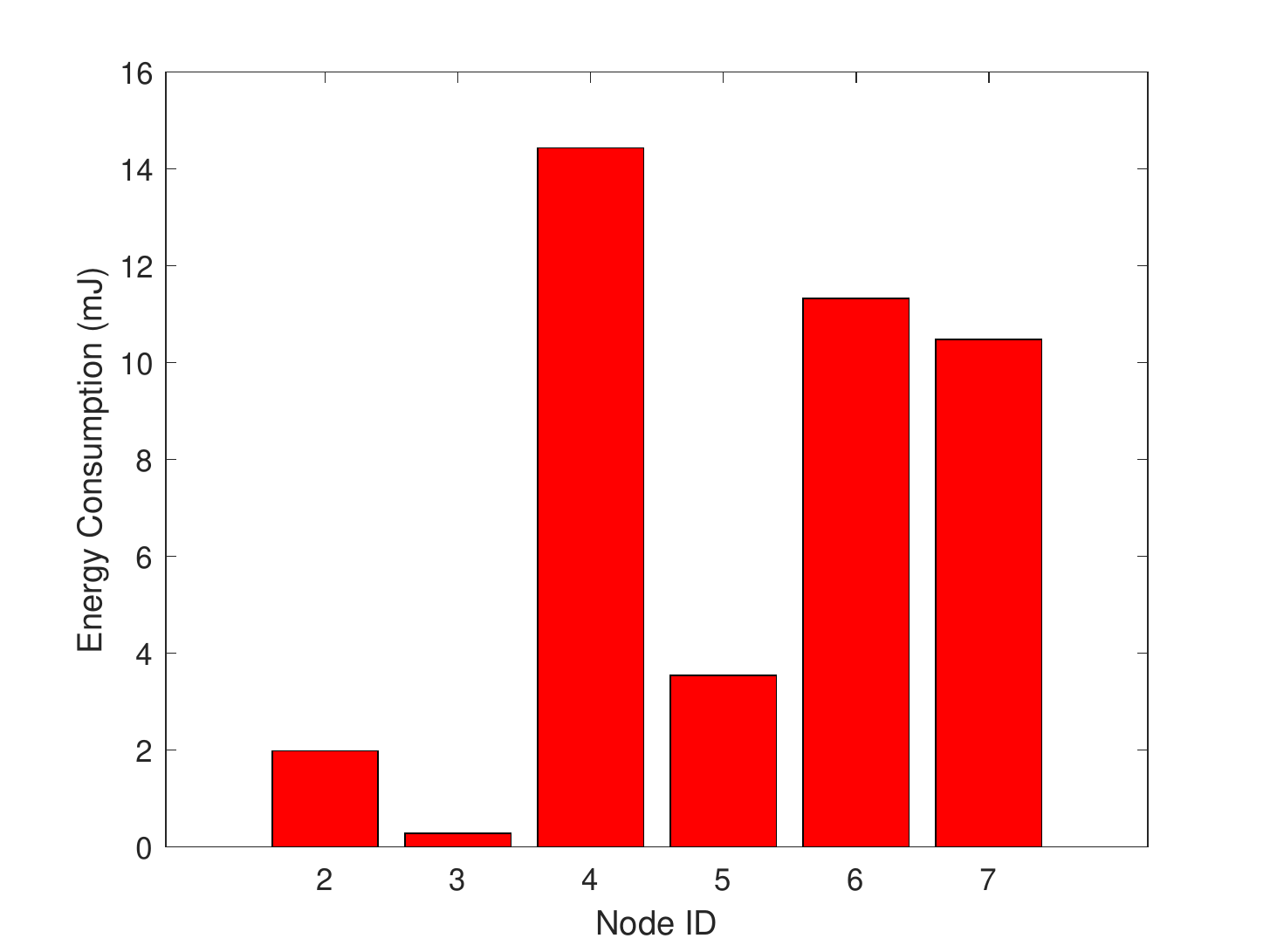}}
     \hfil
     \subfigure[Average backlog or throughput.]
     {\includegraphics[width=0.5\columnwidth]{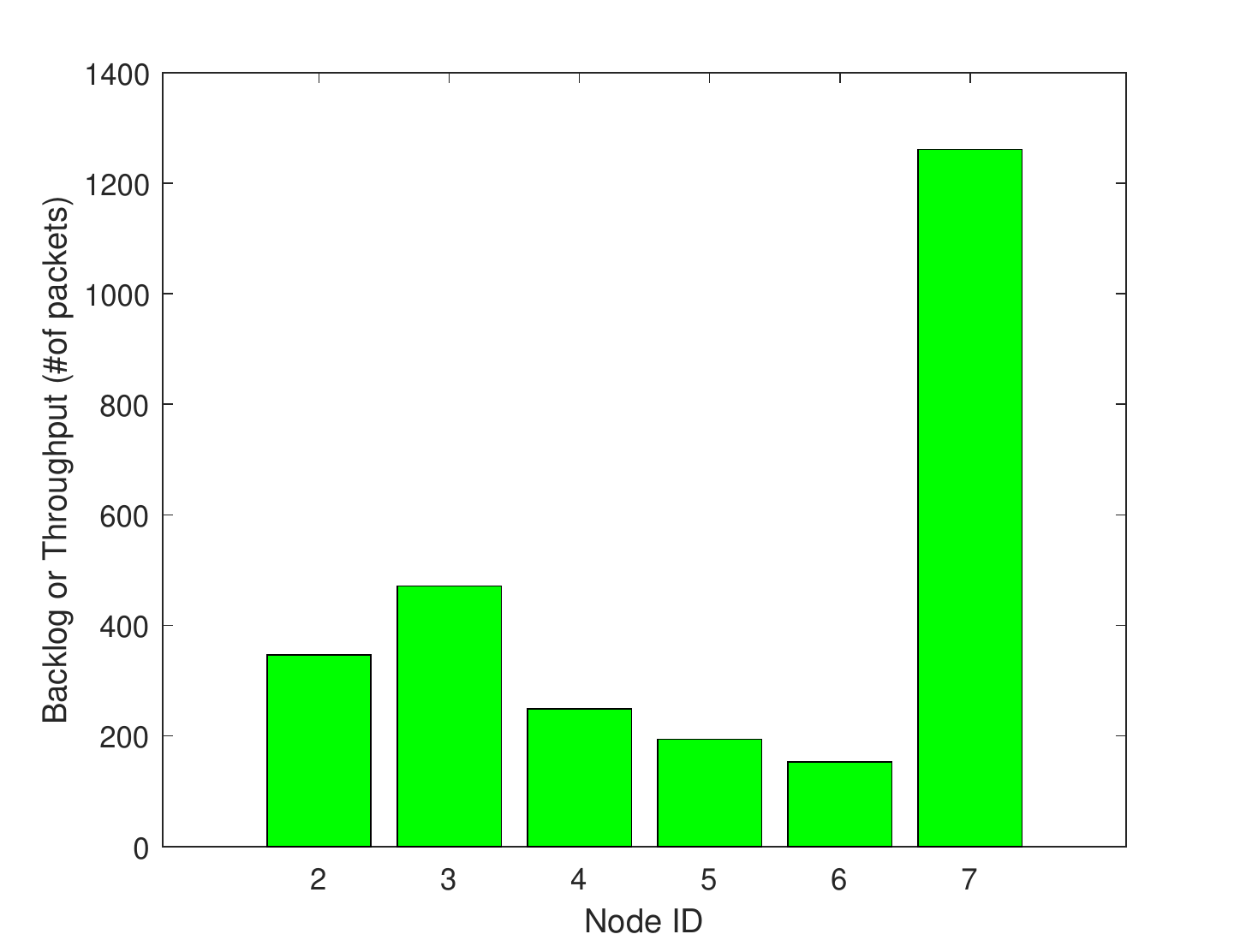}}
     \caption{Performance for ring topology.}
     \label{fig:ring}
 \end{figure*}

The performance results for ring topology are shown in Fig.~\ref{fig:ring}. Nodes 2 and 3 incur higher overhead, since these nodes are the first hop relay nodes for the source node. Since Node 2 has better link from source node compared to Node 3, it is scheduled to transmit more and therefore its overhead is higher to support these transmissions. On the other hand, Node 4 incurs the lowest overhead, since it is a middle relay node on the longer path and therefore on average it receives fewer packets to relay. The energy consumption for Node $4$ is higher than other nodes, since most data packets are relayed through this node. We observe that the destination (Node 7) has relatively higher throughput than the backlogs at relay nodes, since this topology offers two routes for the source to deliver its packets to the destination. Therefore if one route is backlogged, then the backpressure algorithm switches to the other route. The backlogs at the relay nodes show that the algorithm switches periodically between two routes to avoid network congestion.

\begin{figure*}
     \centering
     \subfigure[Average overhead.]
     {\includegraphics[width=0.5\columnwidth]{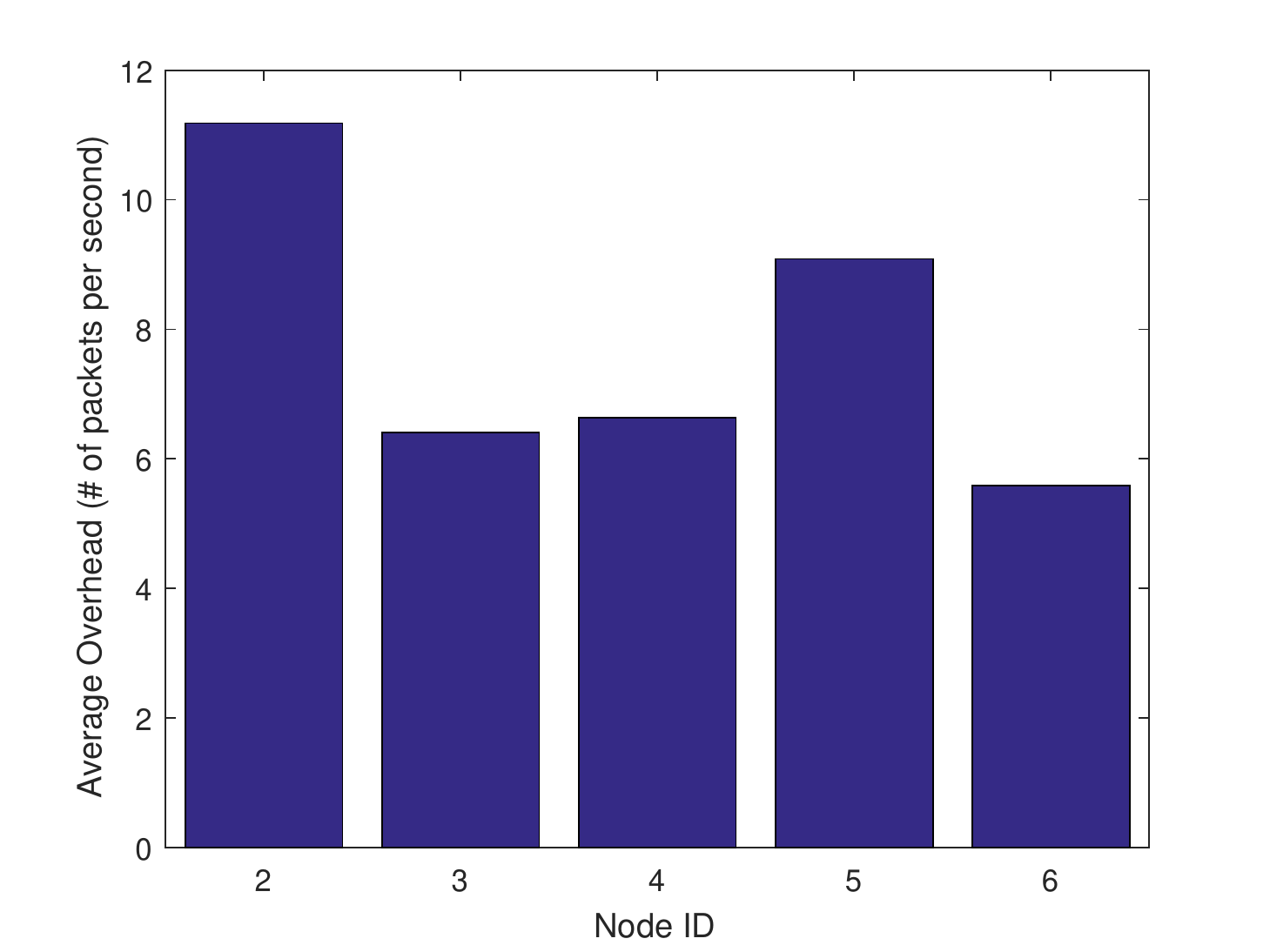}}
     \hfil
     \subfigure[Average energy consumption.]
     {\includegraphics[width=0.5\columnwidth]{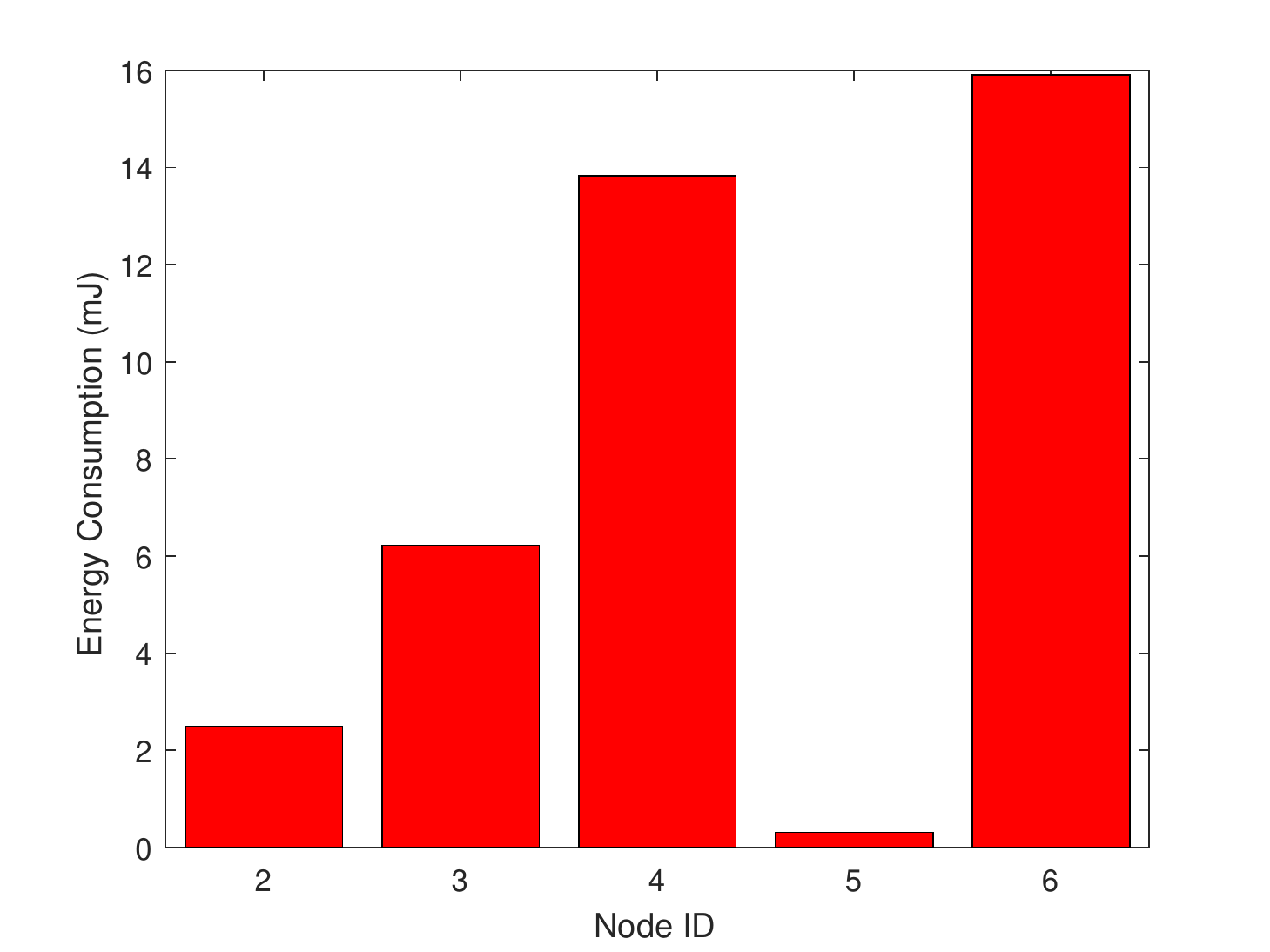}}
     \hfil
     \subfigure[Average backlog or throughput.]
     {\includegraphics[width=0.5\columnwidth]{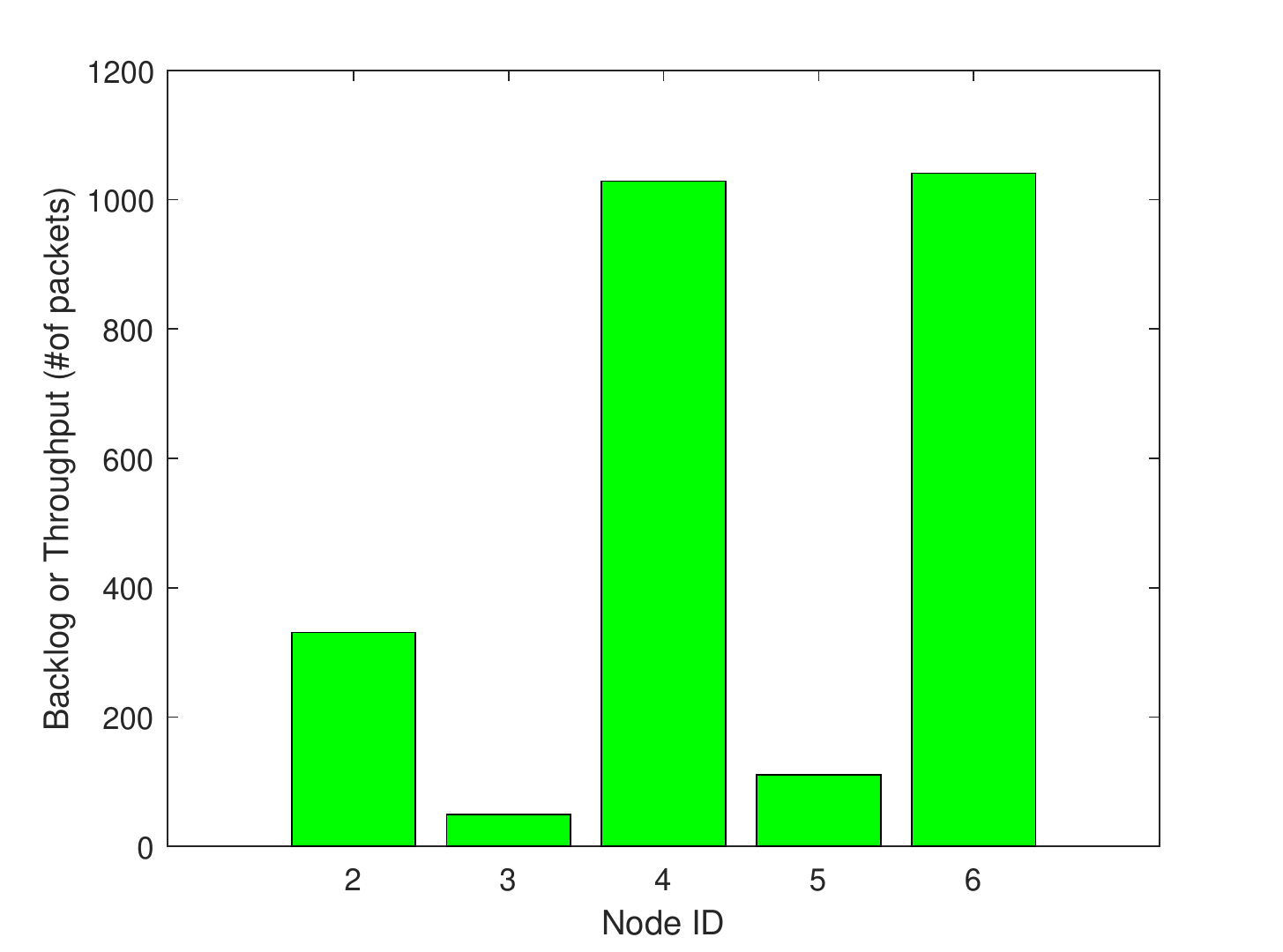}}
     	\vspace{-0.2cm}
     \caption{Performance for grid topology.}
     \label{fig:grid}
 \end{figure*}

The performance results for grid topology are shown in Fig.~\ref{fig:grid}. Node 2  incurs higher overhead than other nodes, since it is the first relay hop from the source to the destination node. The energy consumption of Nodes 3 and 4 is higher than other relays, since most data packets are relayed through these nodes. Node 2 and Node 4 have larger backlogs, since they are the first relay hops from the source to the destination. Node 3 and Node 5 have smaller backlogs, as they can successfully relay (relatively fewer) packets to the destination (Node 6).

\subsection{Emulation Testbed Results with Network Coding}
\begin{figure}
 	\centering
 	\includegraphics[width=0.5\columnwidth]{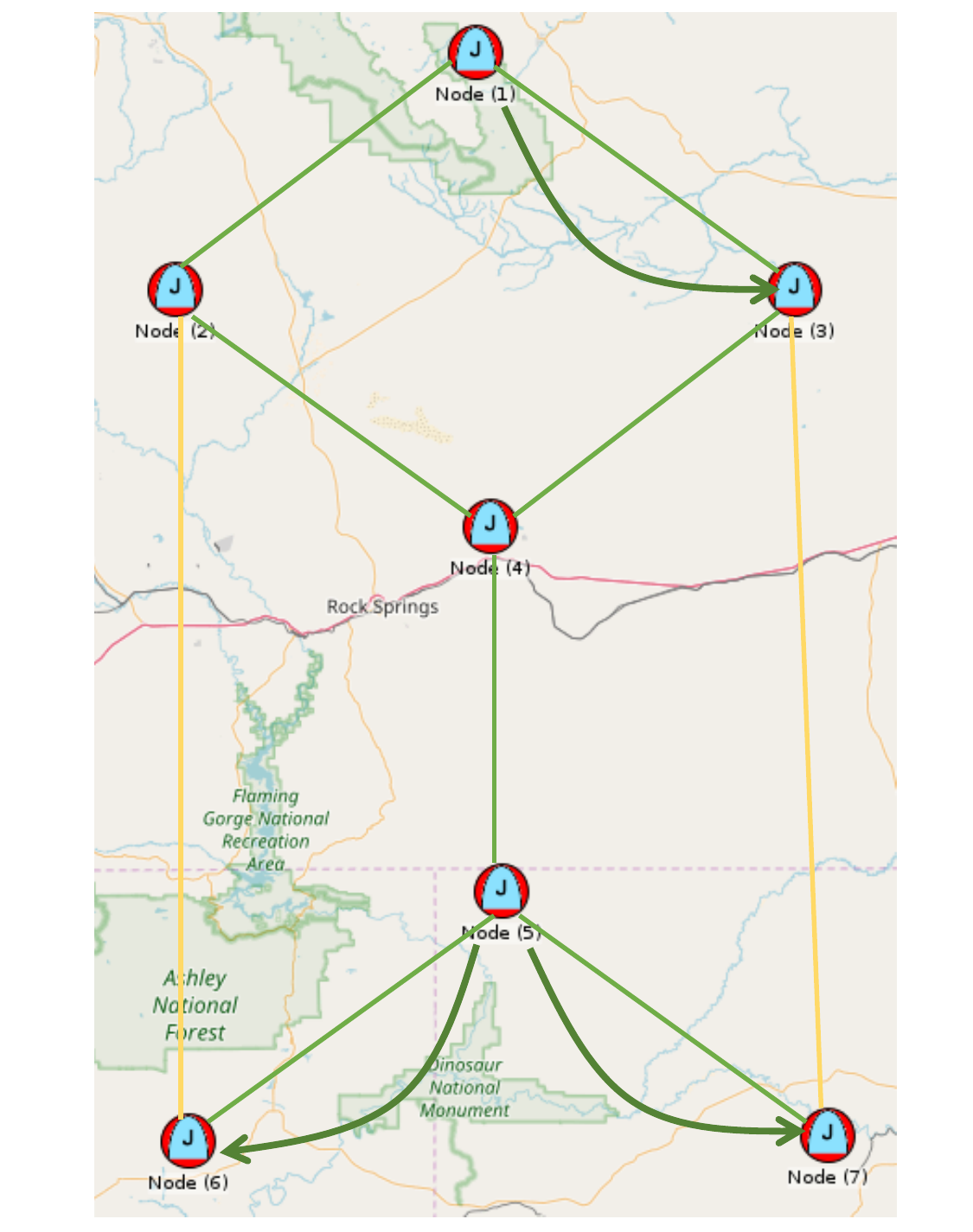}
 	\caption{Topology for emulation tests with network coding.}\label{fig:nc}
 \end{figure}

The scenario (``butterfly") shown in Fig.~\ref{fig:nc} has source node 1 and destination nodes 6 and 7, where a packet contributes to throughput if it is decoded by both destinations.

\subsubsection{Selection of Field Size}
 First, we determine a suitable field size. Table~\ref{table:exectime} shows the execution time (measured on 2.67GHz processor with 8GB RAM) for different field sizes under the same block size and packet length (N/A means the program is not terminated). The use of high field sizes is found computationally inefficient. Therefore, each packet is coded with Galois field GF($2^4$).

 \begin{table}
 	\centering
 	\caption{Execution Time for Different Field Sizes.}
 	\small
 	\begin{tabular}{cc}
 \hline\noalign{\smallskip}
 		Field Size & Execution Time \\
 \noalign{\smallskip}\hline\noalign{\smallskip}
 		GF($2$) and GF($2^2$) & $<$ 1 sec \\
 		GF($2^3$) & 4 secs \\
 		GF($2^4$) & 4 mins 2 secs \\
 		GF($2^5$), GF($2^6$), GF($2^7$), and GF($2^8$) & N/A \\
 \noalign{\smallskip}\hline
 	\end{tabular}
 	\label{table:exectime}
 \end{table}

\subsubsection{Gaussian Elimination}
\begin{figure}
 	\centering
 \includegraphics[width=0.8\linewidth]{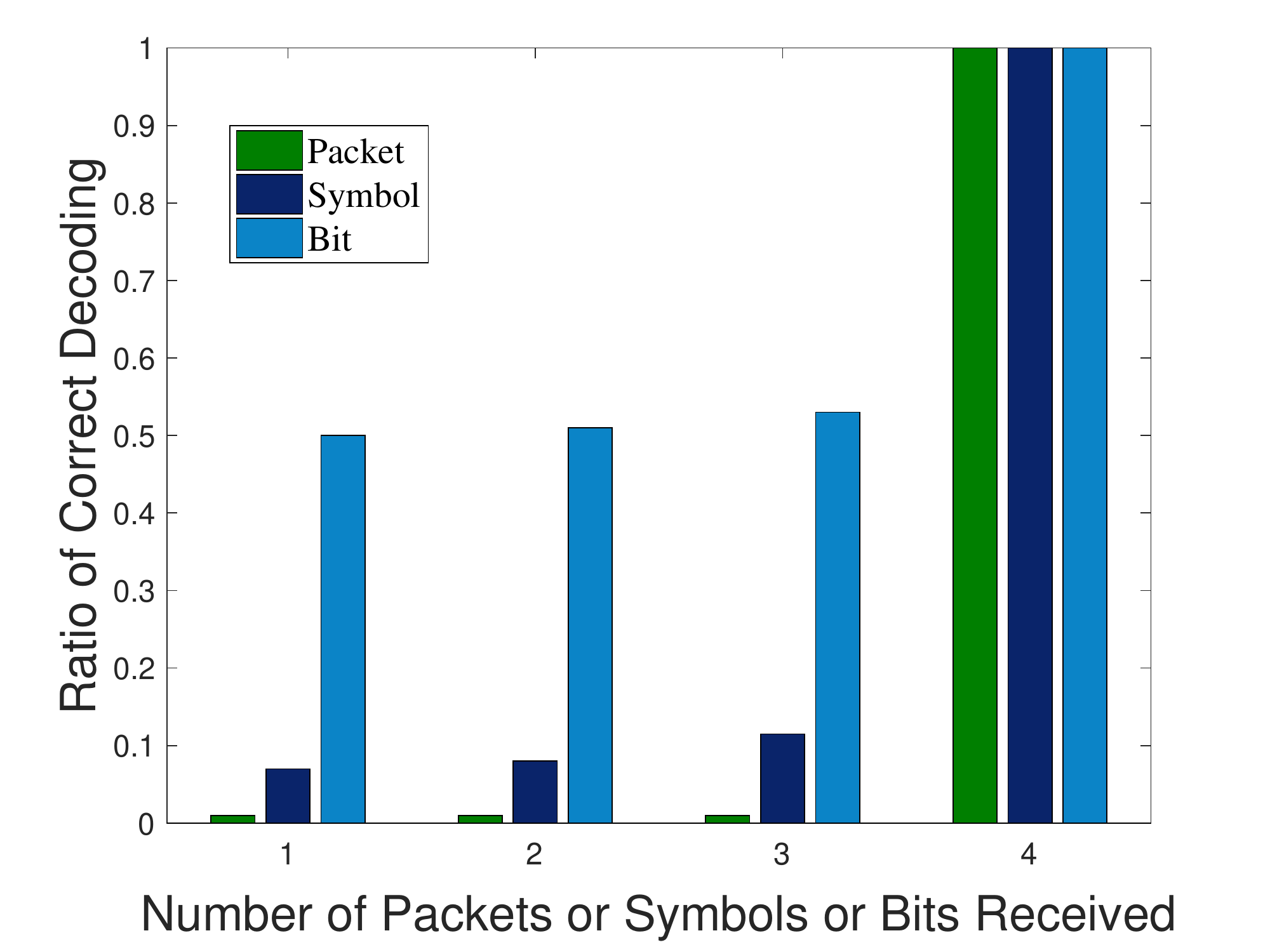}
 \caption{Performance of Gaussian elimination.}\label{fig:succ2new}
 \end{figure}

Gaussian elimination serves as the baseline to decode packets. Note that full rank is not necessary to start Gaussian elimination.
If there is enough information about an individual packet position from the already received packets, then this packet can be successfully decoded. The packets used for a particular linear combination specify a row of the coding matrix. The number of these packets determines how early prior to full rank a packet may be decoded. If coding matrix is highly sparse, each coded packet is derived from fewer packets and therefore fewer packets are needed to decode packets before full rank. The success of the Gaussian elimination decoding is improved as the sparsity of the coding matrix increases (i.e., the coding matrix is closer to an identity matrix). Packets are not decoded correctly when a coded packet is dependent on other coded packets that have not yet been received. A packet is decoded properly when there is enough information and is more likely if the coded packet is dependent on fewer packets.  Fig.~\ref{fig:succ2new} shows the ratio of correctly received packets
as a function of the number of packets
received when Gaussian elimination is used by the destinations.

\subsubsection{Rank Deficient Decoding}
Next, we use \emph{rank deficient decoding} at each destination to decode packets without waiting to accumulate a full rank matrix. The first step is to apply Gaussian elimination to simplify the received data into a form required by the rank deficient decoder. The second step is to use linear programming (LP) decoder to return the lowest weight codeword by leveraging sparsity.

 \begin{table}
 	\centering
 	\caption{Equivalent packets after preconditioning.}
 	\small
 	\begin{tabular}{ccc}
 \hline\noalign{\smallskip}
 		Packets & Before Reordering & After Reordering \\
 \noalign{\smallskip}\hline\noalign{\smallskip}
 		1 & 93.67\% & 99.69\% \\
 		2 & 93.66\%	& 99.59\% \\
 		3 &	93.72\%	& 99.67\% \\
 		4 & 100\% & 100\% \\
 \noalign{\smallskip}\hline
 	\end{tabular}
 	\label{table:precond}
 \end{table}

In the ideal (unrealistic) case, the original data would be used to create a simplified, reduced version of a coded packet, while  the destinations do not have access to the original data in the actual case. In order to get the unreduced, coded data to the same form required by the rank deficient decoding algorithm, we implemented Gaussian elimination prior to using the rank deficient decoding algorithm. It is unknown whether the pre-reduced packets would be equivalent to the normal coded packets after Gaussian elimination in the rank deficient case which could result in error if they were different. To test the validity of this preconditioning step to ensure the input to the decoder is equivalent to the ideal case, a complete block of regularly coded packets is generated and then the partial results of both the simplified case (multiplied by the original data at the decoder) and the regular rank deficient case are compared.
After the full random matrix is generated and a block of coded packets is produced, the simplified version needs to be produced from the same random matrix. For each packet up to the block size, the rows of the random matrix are reduced and then multiplied by the original data to produce simplified versions of the packets. To test equivalency between both cases,  a subset of the normally coded block (packet 1, packets \{1,2\},..., packets \{1, ..., block size\}) is taken and  Gaussian elimination is performed each time. If the normally coded packets are equal to their corresponding simplified versions, then the input to the decoder is correct. For this test, we generated 10,000 different blocks using uniformly distributed random data and random coding matrices with a block size of 4 packets.

The initial results were correct about 93\% of the time. In the failure cases, Gaussian elimination produces a row of zeros which means that there is not enough information. The way the encoding is done, each new packet is guaranteed to be independent and increase the current rank, so transmitting partial data was causing a problem. In cases where the random coding matrix specified the first row has a zero coefficient, the first row is decoded as all zeros without more information available. In order to handle these cases, the random matrix needs to be reordered to be decoded correctly by the receiver. After changing the column order of the random matrix, the results improves to over 99\% accuracy. If the reordering is performed by the decoder instead of by the encoder, the results may be 100\% correct. Taking column permutations by the encoder results in much simpler decoding, however, since the first packet corresponds to the first data packet instead of being out of order. Additionally, taking both row and column permutations in the correct order on the encoding side may produce completely equivalent results, although row permutations degrade the performance when tested. A comparison between unordered vs. reordered packets is shown in Table~\ref{table:precond}. These tests show a high rate of success producing the correct input which shows that  rank deficient decoding results are dependent on the rank deficient decoding algorithm and are not caused by providing incorrect input to the decoder.

\begin{figure}
 	\centering
 \includegraphics[width=0.8\linewidth]{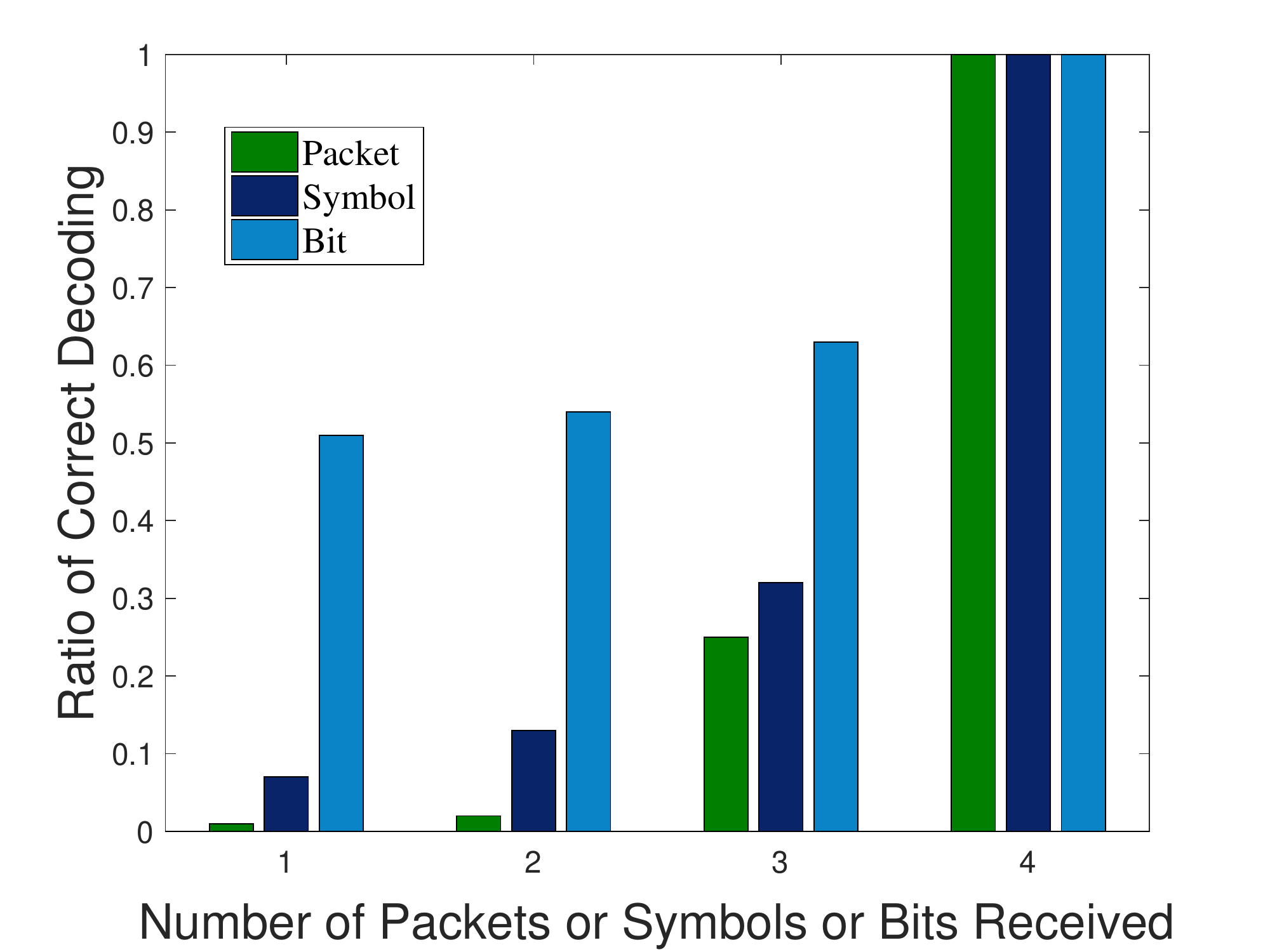}
 \caption{Performance of rank deficient decoding.}\label{fig:succnew}
 \end{figure}

Fig.~\ref{fig:succnew} shows the ratio of correctly received packets
as a function of the number of packets
received with rank-deficient decoding. In this case, 65\% of bits can be correctly decoded before waiting to accumulate the full rank matrix, thereby demonstrating the potential to reduce the decoding delay.

\subsubsection{Throughput, Energy Consumption and Backlog Performance}
\begin{figure}
 	\centering
 	\includegraphics[width=0.8\columnwidth]{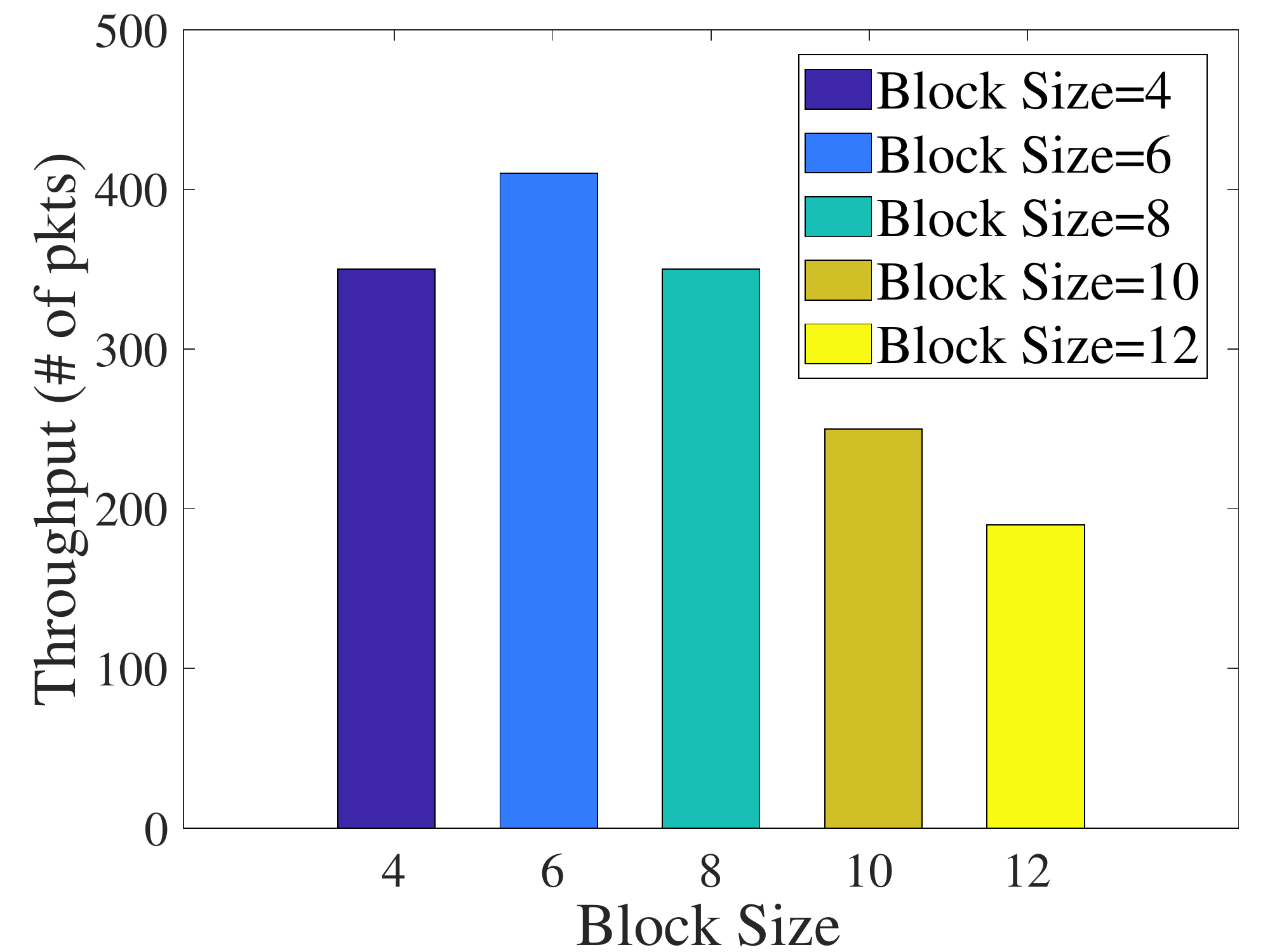}
 	\caption{Throughput for different network coding block sizes.}
 \label{fig:throughput}
 \end{figure}

A detailed end-to-end evaluation of network coding performance is presented next in terms of throughput, energy, and backlog. Fig.~\ref{fig:throughput} shows the impact of \emph{network coding block size} on the \emph{throughput} (the number of packets successfully decoded at the multicast destinations during the course of emulation). The block code size of six (packets) achieves the highest throughput.
However, the throughput drops for higher block
code size due to high overhead in coding.

 \begin{figure}
 	\centering
 \includegraphics[width=0.8\columnwidth]{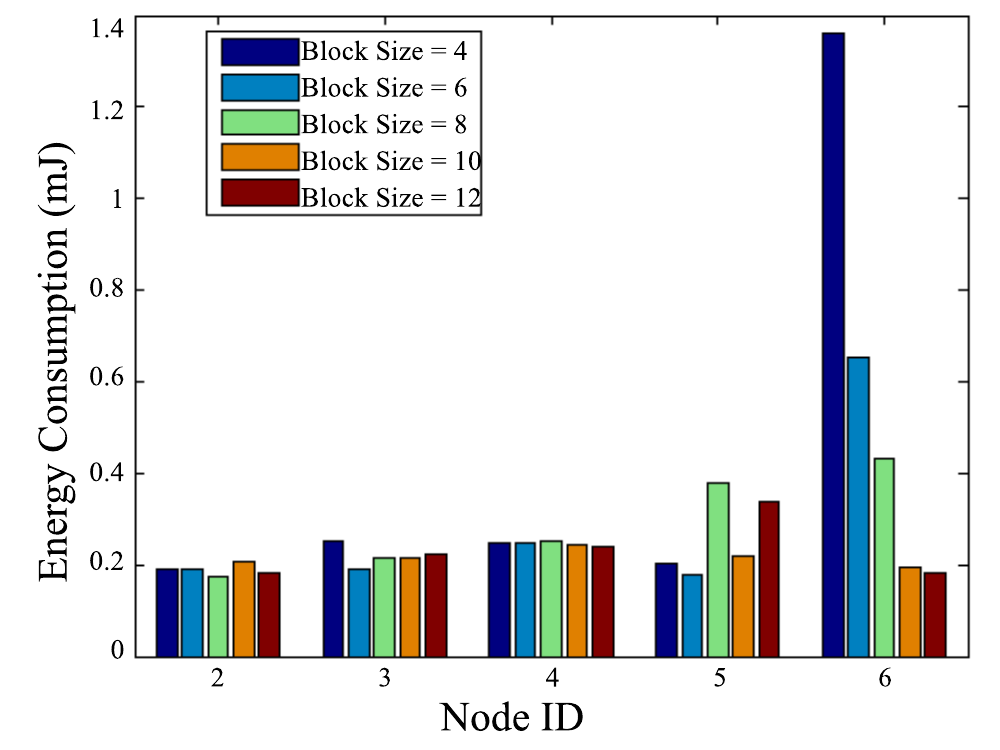}
 \caption{Energy consumption vs. network coding block size.}\label{fig:energy}
 \end{figure}

 \begin{figure}
 	\centering
 	\includegraphics[width=0.8\columnwidth]{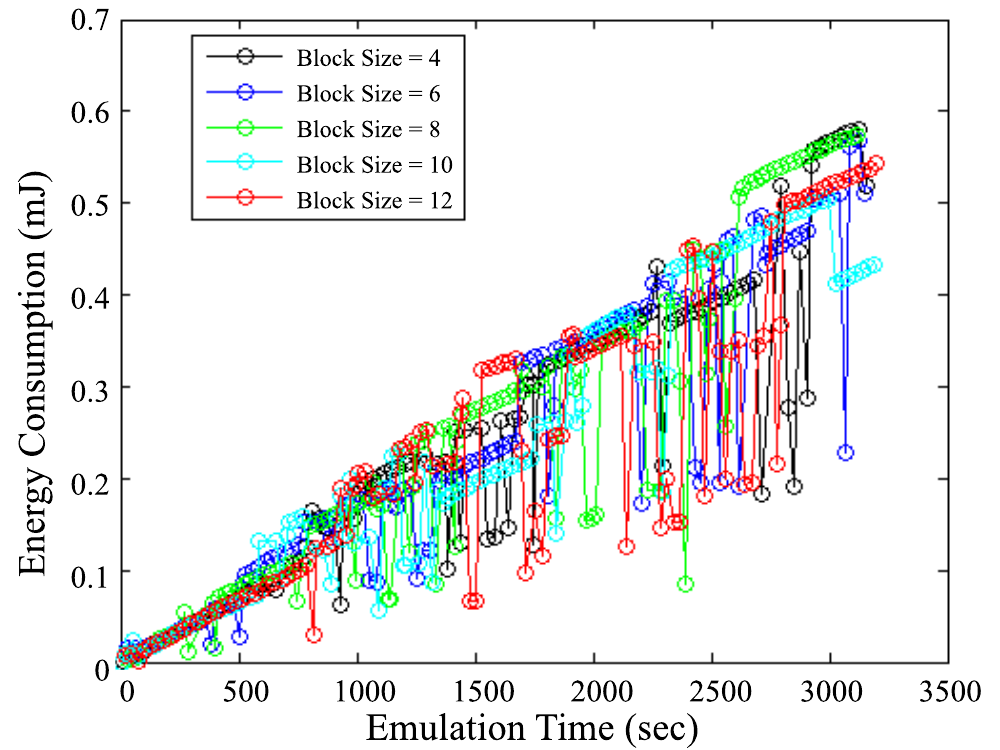}
 \caption{Total energy consumption over time.}\label{fig:8}
 \end{figure}

Next, we measure the \emph{energy consumption}.
Fig.~\ref{fig:energy} shows that the energy consumption (in mJ) at different nodes can be significantly different, where Nodes 6 and 7 consume most energy. For Nodes 2, 3, and 4, the block size does not affect the energy consumption much while for Nodes 5, 6 and 7, different block sizes yield significantly different energy consumptions. The energy consumptions of Nodes 6 and 7 are significantly higher when block size is four. This represents a situation when a node fails to find an available channel to communicate with its neighbors and has to switch channels and retransmit its queue updates to find and establish a connection with its neighbor. Fig.~\ref{fig:8} shows the energy consumption variations of nodes over the course of the emulation test. The energy consumption of nodes increases (close to linear on average) as nodes spend more time in the network.

\begin{figure}
 	\centering
 		\includegraphics[width=0.8\columnwidth]{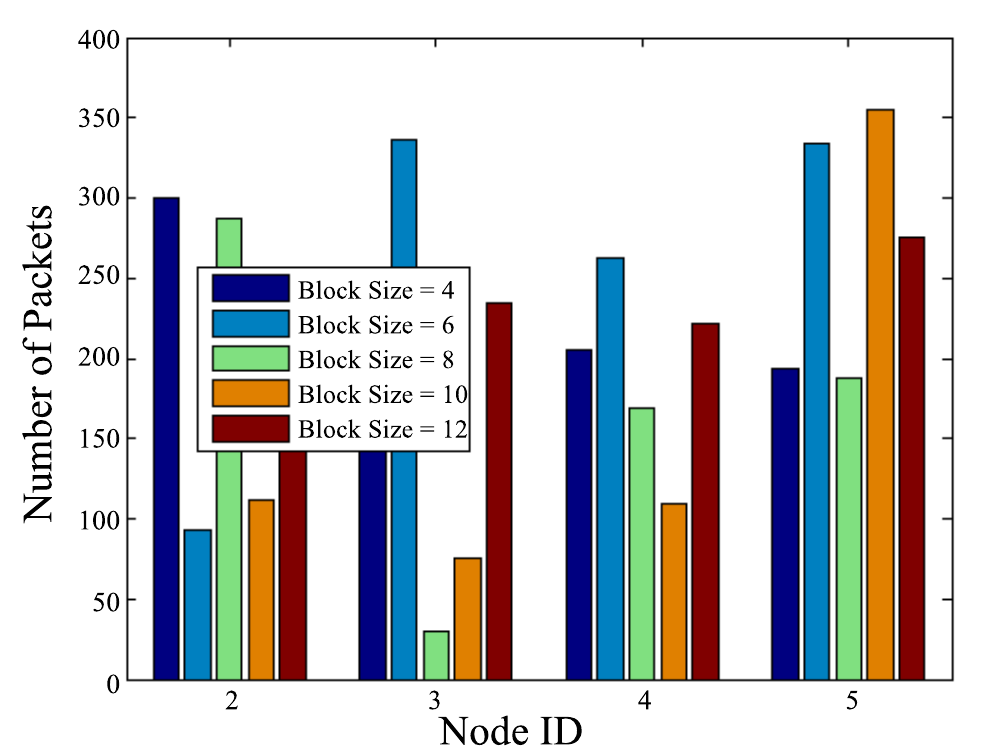}
 		\caption{Average backlog for different network coding block sizes.}\label{fig:9}
 \end{figure}

\begin{figure}
 	\centering
 		\includegraphics[width=0.8\columnwidth]{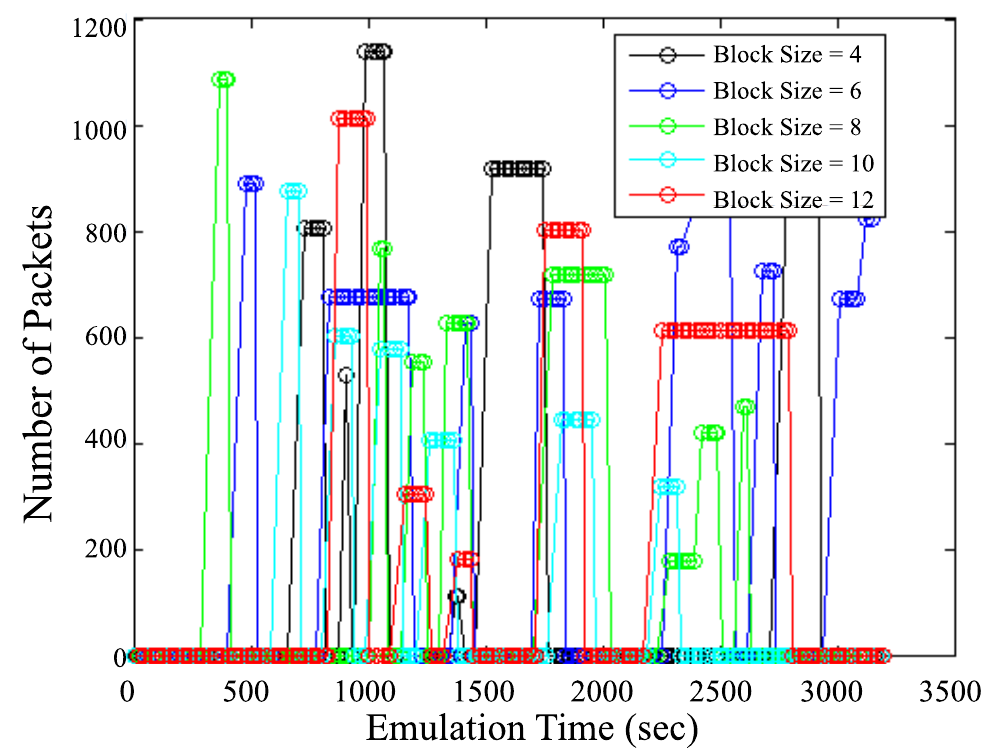}
 		\caption{Backlog over time.}\label{fig:10}
 \end{figure}

Finally, we measure the queue \emph{backlog} at relay nodes (nodes 2-5). Fig.~\ref{fig:9} shows that the effect of block size varies at different nodes. Node 2 has the largest backlog when block size is four, while Node 3 has the largest backlog when block size is six. Nodes 4 and 5 have relatively higher average backlogs in comparison with nodes 2 and 3, since nodes 4 and 5 act as a bridge to destination nodes 6 and 7 in the butterfly topology and more packets travel through node 4 and 5 on the average.
Node 5 has to distribute coded packets to both nodes 6 and 7 for successful network coding.
Hence, Node 5 has the largest backlog in comparison with other relay nodes. Fig.~\ref{fig:10} shows the backlog variations of relay Nodes 2, 3, 4 and 5 over time.

\section{Conclusion} \label{sec:con}

We designed a practical solution for joint DSA, backpressure algorithm and network coding to support both unicast and multicast traffic.
We implemented this solution on SDRs and evaluated its performance in a network emulation testbed. Each SDR runs the identical code for a full protocol stack with distributed coordination (no common control channel) and relies on local information exchange only. Spectrum opportunities are discovered and efficiently utilized by each SDR that executes four phases of distributed coordination to detect and utilize: 1) neighborhood discovery and channel estimation; 2) exchange of flow information updates and execution of backpressure algorithm; 3) transmission decision negotiation; 4) data transmission. We also extended the backpressure algorithm to multicast traffic by implementing network coding with virtual queues for different flows per session and destination. Then we applied full rank and rank deficient decoding to decode network-coded packets. We implemented the cognitive radio capabilities as GNU Radio and Python modules running with USRP N210 as RF front ends.
Then we set up a high fidelity T\&E system and demonstrated the performance in a configurable emulation testbed  with USRP N210 radios communicating over emulated channels according to the cross-layer protocol stack.

\section*{Acknowledgments}
We thank Dr. Bruce Suter and Dr. Zhiyuan Yan for valuable comments and initial code for rank deficient decoding.

\end{document}